# Modeling of technological performance trends using design theory

Subarna Basnet
Massachusetts Institute of Technology, Department of Mechanical Engineering,
77 Massachusetts Ave, Cambridge, Massachusetts 02139

Christopher L. Magee
Massachusetts Institute of Technology, Institute for Data, Systems, and Society,
77 Massachusetts Ave, Cambridge, Massachusetts 02139

## Abstract

Functional technical performance usually follows an exponential dependence on time but the rate of change (the exponent) varies greatly among technological domains. This paper presents a simple model that provides an explanatory foundation for these phenomena based upon the inventive design process.

The model assumes that invention - novel and useful design- arises through probabilistic analogical transfers that combine existing knowledge by combining existing individual operational ideas to arrive at new individual operating ideas. The continuing production of individual operating ideas relies upon injection of new *basic* individual operating ideas that occurs through coupling of science and technology simulations.

The individual operational ideas that result from this process are then modeled as being assimilated in components of artifacts characteristic of a technological domain. According to the model, two effects (differences in interactions among components for different domains and differences in scaling laws for different domains) account for the differences found in improvement rates among domains whereas the analogical transfer process is the source of the exponential behavior. The model is supported by a number of known empirical facts: further empirical research is suggested to independently assess further predictions made by the model.

# Nomenclature and terminology

$Q_J$ = intensive performance of artifacts within a technological domain, J
t = time
IOI = individual operating ideas
$P_{IOI}$ = probability of combination of any two IOI
$IOI_0$ = basic IOI - IOI that first introduce a natural phenomenon in the Operations regime
$IOI_C$ = cumulative number of IOI in the Operations regime
$IOI_L$ = maximum number of possible IOI in Operations regime at time t
$IOI_{SC}$ = $IOI_C$ successfully integrated into a domain artifact
K = annual rate of increase in $IOI_c$ in the Operations regime
$K_J$ = annual rate (when time is in years) of performance improvement measured by the slope of a plot of ln $Q_J$ vs. time
$f_i$ = fitness in Understanding regime for a scientific field i
$F_U$ = cumulative fitness of Understanding regime
$d_J$ = interaction parameter of technological domain J defined as interactive out-links from a typical component to other components in artifacts in domain J
$s_J$ = design parameter affecting the performance of an artifact in domain J
$A_J$ = exponent of design parameter in power law for domain J, relating performance and the design parameter

# 1. Introduction

Inventions are the outputs of the design process when they reach sufficient novelty and utility to rate that term: they are a basic building block of technological progress and the fundamental unit of this paper. In our formulation, *technological domains* consist of *designed artifacts that utilize* a *specified body of knowledge* to achieve a *specific generic function* (Magee et. al. 2014). Thus, technological domains involve a large number of inter-related inventions as even single artifacts can embody multiple inventions. Arthur (2006) used the term "technologies" to describe something that bridges inventions and technological domains; according to Arthur, these use "effects" to achieve some "purpose". Thus, one can also say that each artifact is a material realization of its design that *intentionally* embodies the effects.

This paper brings together three bodies of research that do not usually interact. The first is the design research field, particularly its cognitive scientific insights on the design process. The second is the technological change field where most researchers have been economists or business scholars. The third area is quantitative modeling of performance of artifacts.

The objective of the work reported here is to use understanding of the design and invention process to model performance - how well a specific designed artifact achieves its intended function or purpose. In particular, we examine *performance trends* - the time dependence of performance as realized in *a series of improved designs* of artifacts that arise over time. We do so in an attempt to develop an explanatory and quantitative predictive model for why performance improves exponentially over multiple designs with widely varying rates among technological domains, ranging from 3 to 65% annually for domains characterized so far. Our research question is whether a quantitative predictive model based upon foundations and insights about the design process leads to results consistent with this exponential behavior and whether such a model helps explain and possibly predict the variation in the rate of improvement. We first discuss some relevant literature in each of the three intersecting fields.

## 2 Background

### 2.1 Design, invention and cognitive psychology literature

What connections between technological change and design research can be inferred from the existing literature? Business scholars and economists often view technical change as occurring inside a black box, and have usually avoided examining design activities that are the source of technological change. An important recent publication that begins to build a bridge between aspects of design research and the economics of technological change is the paper by Baldwin and Clark (2006). These authors (and Luo et al. 2014) point specifically to a central role for design in achieving economic value. In addition to economic perspectives, another view that somewhat ignores design is the linear model accredited to Vannevar Bush (Bush, 1945), which considers technological change occurring through application of science. As a counterview, in his seminal book, *The Sciences of the Artificial*, Herbert Simon (1969, 1996) was the first to highlight that design is an activity standing on its own right, like natural sciences, and has its own set of logic, concepts, and principles. While the primary goal of natural science is to produce predictive explanations of natural phenomena, the primary goal of design is to create artifacts. The design activity is central to creation and improvement of artifacts in all technological domains and involves cognitive activities such as the use of knowledge, reasoning, and understanding. These indisputable cognitive activities have been noted by many scholars who have studied invention and design (Simon 1969, Dasgupta 1996, Gero and Kannengiesser (2004), Hatchuel and Weil 2009).

In the context of realizing higher performance from subsequent generations of artifacts, the role of invention, as one outcome of the design process, is a critical one since improvement in performance of artifacts must strongly reflect the inventions. As Vincenti (1990, pg. 230) puts it, inventive activity is a source of new operational principles, and normal configurations that underlie future normal or radical designs. The operational principles (Polyani 1962, Vincenti 1990) of an artifact describe how its components fulfill their special functions in combining to an overall operation to achieve the function of the artifact.

Models found useful in describing the creative design process include the Geneplore model (Finke, Ward and Smith 1996), topological structures (Braha and Reich 2003), FBS theory (Gero and Kannengiesser 2004), CK theory (Hatchuel and Weil 2009), infused design (Shai et al. 2009), analytical product design (Frischknecht et al. 2009), and other modeling approaches. Although all of these frameworks include – to some degree - the key idea of combining existing ideas (for example, in the form of conceptual synthesis, and blending of mental models described in discussion of the Geneplore model), the framework found most helpful in our modeling of performance changes resulting from a cumulative design process is analogical transfer. Although this idea can be traced as beginning with Polya (1945) or earlier, the framework remains an active area in design research (Clement et al. 1994, Holyoak and Thagard 1995, Goel 1997, Gentner and Markman 1997, Leclerq and Heylighen 2002, Dahl and Moreau 2002, Christensen and Schunn 2007, Linsey et al. 2008, Tseng et al. 2008, Linsey et al. 2012, Fu et al. 2013). Scholars of analogical transfer (Gentner and Markman 1997, Holyoak et al. 1995 and Weisberg 2006) explain analogical transfer as involving the use of conceptual knowledge from a familiar domain (base) and applying it to create knowledge in a domain with similar structure (target): analogical transfer exploits past knowledge in both the base and target domains. The analogies utilized can be local, regional or remote, depending on surface and structural similarities between objects involved in the base and target domains. Weisberg discusses the example of the Wright brothers using several analogical transfers to first recognize and solve the problem of flight control. First, they viewed flying as being similar to biking in which the rider has to be actively involved in controlling the bike, an application of regional analogy. Interestingly, many others attempting to design artifacts for flying did not access this regional analogy and thus did not even identify the key control problem. Second, the Wright brothers studied birds to see how they controlled themselves during flight, and learned that they adjusted their position about the rolling axis using their wing tips. From this insight, they had the idea of using similar moving surfaces, another instance of using regional analogy. Lastly, they developed the idea of warping the wings, demonstrated by using a twisted cardboard box, to act like vanes of windmills to make the airplane roll. The use of three analogical transfers in combination to see and solve the flight control problem is a clear case of analogical transfer but there is also evidence (cited earlier in this paragraph) of much wider applicability.

There are more abstract versions of combinatorial analogical transfer that have been proposed in the wider literature. Based on an extensive historical study of mechanical inventions and drawing insights from Gestalt psychology, Usher (1954) proposed a cumulative synthesis approach for creation of inventions. The notion of bisociation (Koestler 1964, Dasgupta 1996) develops the cumulative synthesis approach further and says that a new inventive idea is ideated combining disparate ideas. More recently, Fleming (2001) and Arthur (2006) have respectively used the same combinatorial notions of invention in studying technological change. Other research in the technological change literature also discusses a related concept that is usually called "spillover". Rosenberg (1982) showed that such technological spillover greatly impacted the quantity and quality of technological change in the United States in the 20th century – a result supported by Nelson and Winter (1982) and Ruttan (2001). Indeed, a recent paper by Nemet and Johnson (2012) states that "one of the most fundamental concepts in innovation theory is that important inventions involve the transfer of knowledge from one technical area to another". We note that these descriptions do not always make a clear distinction regarding whether the transfer is occurring at the idea level or at the artifact level. They are silent regarding how and from where designers or inventors get their disparate ideas to combine and regarding details about the complexities of transfer and combination.

Analogical transfer of ideas as a broad mechanism and expertise as the foundation of ideas (Weisberg, 2006) provides adequate specificity for modeling science and invention in this paper. Weisberg contends that analogical transfer is utilized in generation of both scientific and technological knowledge. Existing knowledge provides the foundational basis for analogical transfer to occur. A similar argument has been applied to the more abstract notion of combinations. Usher describes a cumulative synthesis approach - a four step social process (perception of the problem, setting the stage, the act of insight, critical revision) - which brings together inventive structures to create new inventions. Ruttan (1959), has argued that Usher's formulation provides a "theory of the social processes by which 'new things' come into existence that is broad enough to encompass the whole range of activities characterized by the terms science, invention, and innovation". Models of both Understanding and Operations regime in our paper (defined in the next paragraph) utilize the abstraction that knowledge is created by probabilistically[1] combining existing knowledge made available by analogical transfer.

Vincenti (1990), and Mokyr (2002) take the view that scientific and technological knowledge can be classified into descriptive (Understanding) and prescriptive (Operations) knowledge[2] regimes. The Understanding regime can be seen as a body of 'what' knowledge and includes scientific principles and explanations, natural regularities, materials properties, and physical constants. A unit of understanding is falsifiable (Popper 1959) and enables explanation and prediction about specific phenomena, including behavior of artifacts. The Operations regime, on the other hand, can be viewed as a body of 'design knowledge', which suggests how to leverage natural 'effects' (Arthur, 2006, Vincenti, 1990)) to achieve a technological advantage or purpose. It includes, operating principles, design methods, experimental methods, and tools (Dasgupta 1996, Vincenti 1990). Based on this distinction, understanding enables generation of operational knowledge, which ultimately contributes towards design of some artifact. However, operations is not entirely based upon existing understanding and in fact innovations in know-how can and often do occur before any understanding of related natural effects is available.

An important aspect of design and invention is the cooperative interaction between Understanding and Operations regimes (Musson, 1972, Musson and Robinson 1989). Using a historical perspective, Mokyr (2002) has carefully observed that a synergistic exchange between the two has been occurring, where each enables the other. The contribution of Understanding to Operations is well known: it provides principles, and regularities of natural effects, including new ones, in the form of predictive equations, and descriptive facts, such as material properties. Fleming and Sorenson (2004) provide evidence that understanding helps inventors by providing a richer map to search for operating ideas, which can be combined together. Understanding also provides insight about where new technological opportunities may be found (Klevoric et al. 1995). Beyond these contributions, there is the more general view, discussed in the initial paragraph of this section, that new operational ideas can be derived from new understanding. What is less discussed is the multi-faceted contributions of Operations to the Understanding regime. In his paper, *Sealing wax and string*, de Solla Price (1983), a physicist, and historian of science, highlighted that instruments (an output of the Operations regime) were a dominant force in enabling scientific revolutions. He states: "changes in paradigm that accompany great and revolutionary changes (in science) were caused more often by application of technology to science, rather than changes from 'putting on a new thinking cap' ". Operations provide tools and instruments to make measurements, and to make new discoveries. In his book, *The Scientist: A History of Science Told Through the Lives of its Greatest Inventors*, Gribbin (2002), a British astrophysicist, and science writer, has described how the ability to grind eyeglass lenses made it possible to make better telescopes, and hence paved the way for astronomers to make new discoveries. New or improved observational techniques are still a major driver of progress in science. Gribbin has aptly summarized the enabling exchange between the two regimes: "new scientific ideas leading… to improved technology and new technology providing scientists with the means to test new ideas to greater and greater accuracy". Additionally, the Operations regime provides new problems for the Understanding regime to study, and has led to birth of new fields in Understanding (Hunt 2010). Based upon these insights and with our focus on explaining performance improvement arising from continuing streams of inventions, our model treats mutual exchange between Understanding and Operations.

---

[1] At a point in time, not all possible combinations of existing knowledge lead to new knowledge.

[2] We use the terms "Understanding" and "Operations", since each one brings more clarity to the nature of underlying activity. Understanding refers to conceptual insight that is generated about an object or environment, whereas Operations refers to the idea of acting on an object or environment to get some desired effect, as well as experimental methods included in the term 'science'.

In design of artifacts, Simon (1962) introduced the notion of interactions in his essay on the complexity of artifacts. When a design of an artifact is changed from one state to another (with differences between the two states as defined by multiple attributes, say D1, D2, and D3) by taking some actions (say, A1, A2, and A3), in many cases, any specific action taken may affect more than one attribute, thus potentially manifesting as interactions of the attributes. The same notion of interaction/conflicts is captured by the concept of coupling of functional requirements (Suh 2001), or dependencies between characteristics (Weber 2003), which can occur when two or more functional requirements are influenced by a design parameter. Theoretically it seems ideal to have one design parameter controlling one functional requirement to achieve a fully decomposable (modular) design (Suh 2001, Baldwin and Clark 2000). However, Whitney (1996, 2004) has argued that, in reality, how decomposable a design of an artifact can be depends on the physics involved or additional constraints, such as permissible mass. These are reflected as component-to-component, and component-to-system interactions, or as a need to have multi-functional components. Consequently, Whitney argues, complex electro-mechanical-optical (CEMO) systems, primarily designed to carry power, cannot be made as decomposable as VLSI systems primarily designed to transmit and transform information. For example, in energy applications, the impedance of transmitting and receiving elements has to be matched for maximum power transfer, thus making the two elements coupled. Further, CEMO systems typically need to have multi-functional components in order to keep the artifact size reasonable, creating coupling of attributes at the component level. Another type of interaction Whitney has identified are the side effects, such as waste heat in computers, and corrosion of electrodes in batteries - that occur in artifacts, which in some electro-mechanical systems can consume significant portion of the design effort for their mitigation. The presence, and thus the resolution, of these different interactions cause significant delay, consume significant engineering resources and potentially stop applications of some concepts, thus making the level of interactions of a technological domain a potentially strong factor influencing its rate of improvement. Based upon Whitney's work, the effect of interactions on rates of improvement was suggested qualitatively by Koh and Magee (2008) and a quantitative model of the effect was developed by McNerney et al. (2011) – see section 2.3.

The influence of design parameters on artifact performance is an essential part of design knowledge. Many technological domains have complex mathematical equations relating some aspects of performance with design parameters. Indeed, the so-called engineering science literature has such equations for many aspects affecting the design of artifacts of perhaps all technological domains. Simpler relationships concerning the geometrical scale of artifacts are also available and generally give performance metrics as a function of a design variable raised to a power. Use of power-law relationships can be found in: 1) Sahal (1985) who studied scaling in three different sets of artifacts - airplanes, tractors, and computers; and 2) Bela Gold (1974) who demonstrated that doubling the size of a blast furnace reduces their cost by about 40%. The constant percent change per doubling in size results from the power law (assumed by Gold) between performance/cost and geometrical variables such as volume.

## 2.2 Technological change literature

What descriptive models and theories help us understand why technologies improve and how the improvement patterns are structured? Schumpeter (1934) introduced the idea that entrepreneurs, whose primary role is to provide improved products and services through innovation, drive economic progress. These innovations, which Schumpeter describes as *industrial mutations*, displace competing products and services from the economy. However, they, too, are displaced by higher performing innovations that follow, thus perpetuating the cycle of *creative destruction*. Building upon Schumpeter's notion, Solow (1956) recognized and incorporated technological change as the key element in his quantitative explanatory theory of economic growth. The basic conclusion that technological change is the foundation of sustained economic growth has stood the test of time. Later theorists of economic growth (Arrow 1962, Romer 1990, Acemoglou 2002) have attempted to deal with the more complex problem of embedding technological change within the economy (endogenous to different degrees). Although the later theories are important, the issues are outside the scope of this paper and will not be covered here. A related question of demand-pull and technology-push does have more relevance.

What drives technological innovation? Some early explanations emphasized pure demand push (Carter and Williams (1957, 1959), Baker et al. 1967, Myers and Marquis 1969, Langrish et al. 1972, Utterback 1974) where the needs of the economy at a given time dictate technological direction. Mowery and Rosenberg (1979) reanalyzed the data and methodology in this early work and arrived at a strong role for science/technology push (the discoveries of scientists and inventors primarily determine technological direction). Taking a balanced view, Dosi (1982) argued that both market-pull (customer needs and potential for profitability) and technology-push (in the form of promising new technology, and the underpinning procedures) are equally important for being sources of innovation.

Tushman and Anderson (1986) discuss discontinuities as having large socio-technical effects and note that such discontinuities are an essential element of technological change. In another highly referenced paper, Henderson and Clark (1990) emphasize the importance of architectural change of artifacts - as opposed to component change - having large effects on the firm-level impact of change. Christensen (1996), on the other hand, views technological change occurring as a series of disruptive product innovations that start in a niche market catering to different functional requirements, but then rapidly improve towards the requirements of mainstream performance. The disruptive technology surpasses the mature market leaders (by achieving the necessary performance in smaller, cheaper artifacts), and displaces them.

All of the concepts of technological change described in the preceding paragraphs - at least implicitly - depend upon relative rates of change of performance. This is the focus of our modeling effort so we will now briefly review concepts related to trends in performance of designed artifacts, and what patterns they have followed. We first review two established frameworks – generalizations of Wright's early research, and Moore's Law - for describing trends in technological performance. In 1936 Theodore Paul Wright (1936) in his seminal paper "Factors affecting the Cost of Airplanes" for the first time introduced the idea of measuring technological progress of artifacts. From his empirical study of airplane manufacturing, he demonstrated that labor cost or total cost of specific airplane designs decreased as a power law against their cumulative production. This relationship is expressed as:

$$C = C_0 P^{-w} \qquad (1)$$

Where $C_0$, and $C$ are unit cost of the first, and subsequent airplanes respectively, and where P and w are cumulative production and its exponent that relates it to unit cost. Wright explains that labor cost reductions are realized as shop floor personnel gain experience with the manufacturing processes, and material usage and have access to better production tools. Since Wright's work, this approach has been used to study production of airplanes and ships during World War II, and extended to private enterprises (Yelle, 2007). It should be noted that Wright did not look at improvement due to new designs, instead he only considered improved manufacturing of a fixed design.

Gordon Moore (1965) presented the second approach - using time as the independent variable and investigating a series of newly designed artifacts- in his seminal paper that describes improvement of integrated circuits. He observed that the number of transistors on a die was doubling roughly every 18 months (modified to 2 years in 1975). This exponential relationship between the number of transistors on a die and time, famously known[3] as Moore's Law, can be mathematically expressed as:

$$Q_J(t) = Q_J(t_0) \exp\{K_J (t-t_0)\} \quad (2)$$

Where $Q_J(t_0)$ and $Q_J(t)$ are the number of transistors per die (a measure of performance) at time $t_0$ and time $t$, and $K_J$ is the rate of improvement (annual if time is in years). For integrated circuits, the exponential relationship has held broadly true for five decades. Others (Girifalco 1991, Nordhaus 1996, Koh and Magee (2006, 2008) and Leinhard 2008) utilized this temporal approach to study performance of different technologies, and have demonstrated that many technologies exhibit exponential behavior with time. More recently, Magee et al. (2014) extended the study to 73 different performance metrics in 28 different technology domains. The performance curves have continued to demonstrate exponential behavior, although annual rates vary widely across domains. We note that Moore and all others who used his framework basically compared the performance of different designs over time differentiating the Wright and Moore frameworks. However, it is also possible to use the Wright framework for different designs but only if the amount produced increases exponentially with time (Sahal, 1979, Nagy et al. 2013, Magee et al 2014).

In order to clarify for readers the nature of empirical performance data, we present performance data for two sample domains, magnetic resonance imaging (MRI) and electric motors (Fig. 1a), and a summary of improvement rates for 28 domains (Fig. 1b from Magee et al. 2014). The exponential trend for each domain can be described by equation (2), where $Q_J(t)$ and $Q_J(t_0)$ are the intensive performance of an artifact in domain J at time t and $t_0$, and $K_J$ is the annual rate of improvement of the domain in question.

---

[3] This designation was given to the relationship by Cal Tech professor Carver Mead.

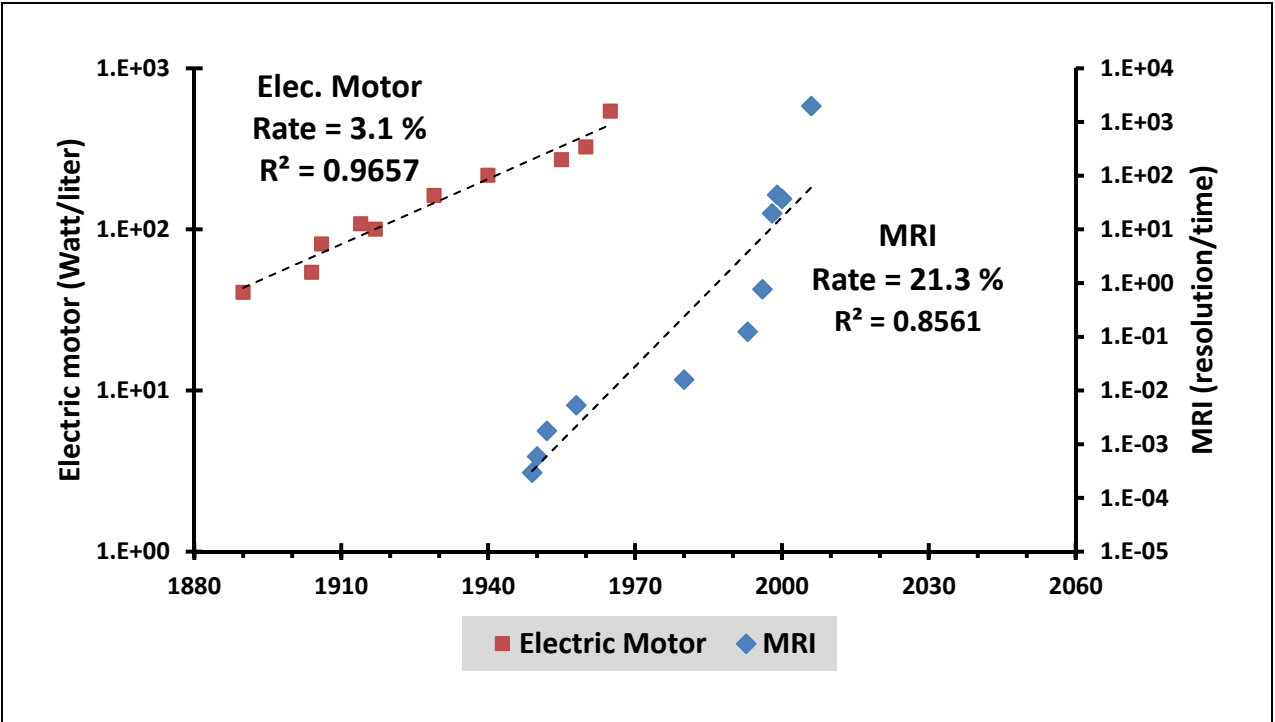

**Fig. 1a:** Exponential growth of performance in sample domains – Electric motor and Magnetic resonance imaging (MRI). Adapted from Magee et al. 2014 with permission.

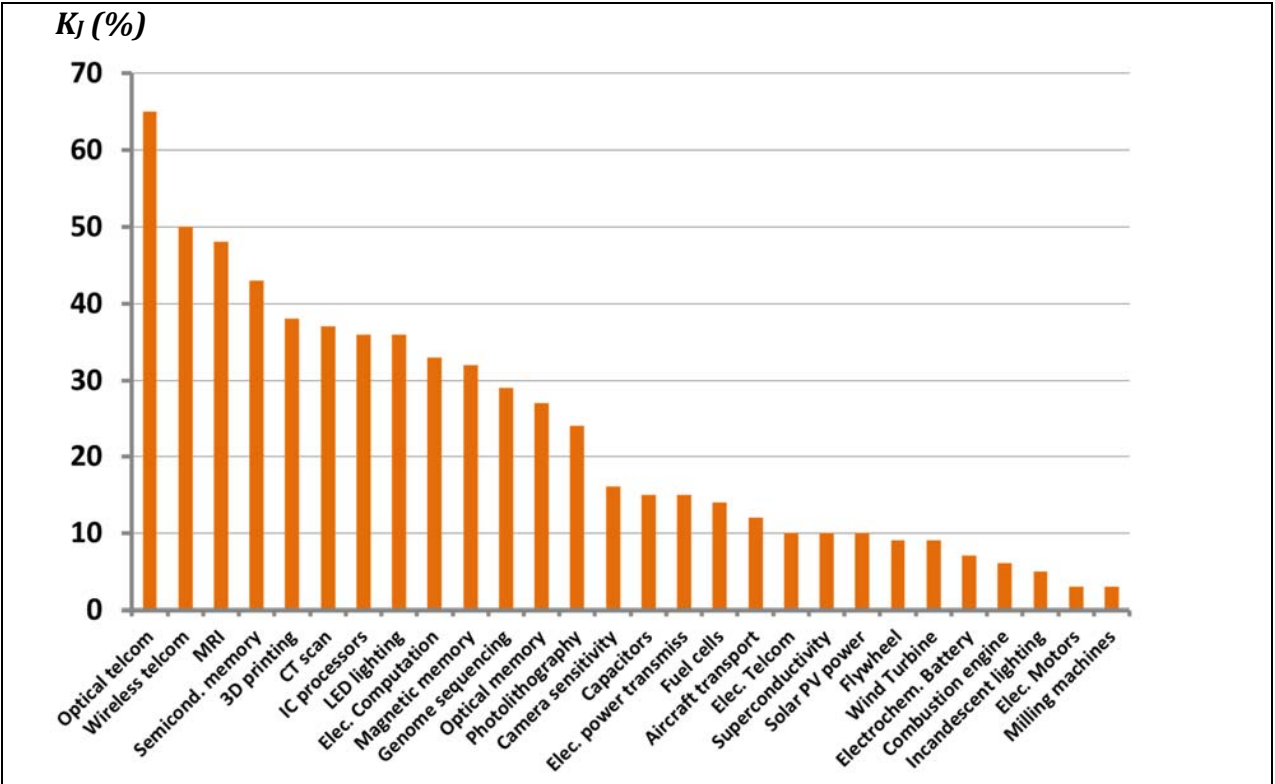

**Fig. 1b:** Annual rate of performance improvement, $K_J$, for 28 domains. Adapted from Magee et al. 2014 with permission.

A recent paper (Benson and Magee, 2015a) has empirically investigated the variation of the improvement rates in these 28 domains. The work has important relationships to the current work so we describe it to note the relationships but to also clarify the fundamental differences. Benson and Magee found strong correlations between specific meta-characteristics of the patents in the 28 domains[4] and the improvement rate in the domains. These authors found that patent meta-characteristics reflecting the importance (citations per patent by other patents), recency (age of patents in a domain) and immediacy (the average over time of the usage of current new knowledge in the domain) are all correlated with the improvement rate. They found a particularly strong correlation (r= 0.76, p =2.1x $10^{-6}$) with a metric that combines immediacy and importance (the average number of citations that patents in the domain receive in their first three years). The findings (and associated multiple regressions) are robust over time and with domain selection and are of practical importance in predicting technological progress in domains where performance data is not available (Benson and Magee, 2015). Nonetheless, the conceptual basis for the findings is observed attributes of the inventive output from a technological field (importance, recency and immediacy of a patent set) and not the process of invention, design knowledge or other technical aspects of designed artifacts in the domain. The aim of the work reported in the present paper is to develop a model that yields insights about the pace of change without recourse to concepts based upon observation of the output over time. If fully successful, we would be able to judge the potential for change based only upon the nature of the design knowledge and we might even be able to find new approaches that might achieve technological goals at more rapid improvement rates.

## 2.3 Literature on quantitative modeling of technological change

What research has attempted to model the technological performance trends that we just discussed? Muth (1986) and Auerswald et al. (2000) have developed models to explain Wright's results by introducing the notion of search for technological possibilities. Each paper assumes that random search, a key element of technological problem solving, for a better technique is made within a *fixed* population of possibilities. Considering a case of a single manufacturing process, Muth (1986) developed a model to capture the idea of substituting manufacturing sequences with better ones. He argues that shop personnel improve the process by learning through experience and making random search for new techniques, which enable improvement of processes leading to cost reductions. Muth demonstrated that the notion of fixed possibilities easily leads to fewer and fewer improvements that can be realized and he argues that the data (for fixed designs) shows a leveling off and eventual stoppage as the model suggests. Building on Muth's idea of random search within a set of fixed design possibilities, Auerswald et al. modeled a multi-process system, in which different processes can be combined to create diverse recipes, and for the first time introduced the notion of interactions by allowing adjoining processes to affect each other's cost.

---

[4] The patents are found by a new technique - Benson and Magee 2015b

Following similar reasoning as Muth and Auerswald et al., McNerney et al. (2011) have developed a stochastic model to explain how the cost reduction of a multi-component system is influenced by component interactions, which they refer to as connectivity between components. McNerney et al. operationalized the notion of interactions as out-links representing influence of a component on other components. When a specific component in a domain artifact changes by introducing a new operational idea, the change affects the design of all the components it influences. If the performance of the artifact (influencing and influenced components) as a whole improves, then McNerney et al. consider the interactions to be resolved and the operating idea is considered successful. The McNerney et al. paper demonstrates that artifacts with more interactions improve more slowly than artifacts with less interactions.

Using agent-based modeling, Axtell et al. (2013) have developed a competitive micro-economic model of technological innovation utilizing the notion of technological fitness. Although they do not discuss or cite Moore's law or his work, they have demonstrated that cumulative technological fitness of all agents increases exponentially overtime. This is different from other researchers who have predominantly been focused on Wright's framework. Consistently, Axtell et al. consider new designs and not just process optimization.

Using a simulation approach, Arthur and Polak (2006) have modeled how new generations of artifacts arise by combining currently available artifacts. The artifacts considered are electronic logic gates. New designs (combinations) are more complex logic gates that can then also be combined into even more complex logic gates. In their model, Arthur and Polak specify several design goals towards to which the logic gates evolve. They have demonstrated that designs with higher levels of complexity cannot be attained without realizing design configurations with intermediate levels of complexity, and new designs with higher functionality substitute for current designs with inferior functionality. This model is much richer than other models in representing the artifact part of the design process; however, it does not consider performance improvement, as do the other models. It is also limited to developing pre-specified artifacts and is thus a specific process; consequently, it is not open-ended or general which are characteristics necessary for modeling performance trends for general technological domains.

Although some are more explicit than others, one feature common to all these models is that all utilize the notion of building upon the performance (in the form of cost) or designs of the past, a key feature of cumulative processes included in the model presented here. On the other hand, they do not consider two aspects we believe useful in answering our research question. First, none of them discusses or includes the influential role played by exchange between science and technology. In this paper, we treat the design process and the exchange between science and technology as important elements for understanding the change in performance over time that in turn is essential to understanding technological change. Second, none consider the design process or operating principles and instead look at combinations at the artifact level instead of combination of ideas.

# 3. Overview of the model

## 3.1 Conceptual basis of model

The desired output from the constructed model are performance improvement rates. To agree with known empirical results, performance should increase exponentially with time. We utilize two sets of mechanisms from design to construct the overall model. The first set, which gives rise to exponential trends, includes growth of knowledge - understanding and operations - using combinatorial analogical transfer aided with mutual exchange between the two. The second set, which gives rise to variation in improvement rates, includes component interactions and scaling of design variables. Since the goal of the model is to develop an explanatory and quantitative predictive model, while modeling these mechanisms we have, where necessary, simplified (removed details) and utilized abstraction to keep the model tractable.

The overall architecture of the model is shown in Figure 2. Based on the work of Vincenti (1990) and Mokyr (2002) that we discussed earlier, we classify scientific and technical knowledge into Understanding and Operations regimes. We further split the Operations regime into idea and artifact sub-regimes where non-physical representation of artifacts are in the idea sub-regime. The idea sub-regime, represented as an ideas pool, consists of individual operating ideas (IOI). The IOI (individual operating idea) concept is an abstraction and generalizes the idea of operating principle introduced by Polyani (1962) and includes any ideas, including operating principles, invention claims, design structures, component integration tricks, trade secrets and other design knowledge that lead to performance improvement of artifacts. An IOI is different than a unit of understanding (UOU) which includes scientific principles, and factual information. An example of a unit of understanding (UOU) is the principle of total internal reflection, which describes how a beam of light undergoes reflection inside a dense medium, when the angle of incidence is above a critical value (see Fig. 3). This principle accurately describes a natural effect, but it does not prescribe how we can use it to transmit information. On the other hand, a pair of parallel surfaces (or a fiber) enclosing a dense medium and utilizing the principle of total internal reflection provides a mechanism – an operating principle - to make a ray of light travel down the length of the medium (see Fig. 3). Such a mechanism is an example of an IOI. Unlike artifacts, which belong to a specific technological domain, we model IOI in the ideas (IOI) pool as being non-domain specific and available to all technological domains. For instance, the operating principle of total internal reflection is utilized in fiber optic telecommunications, fluorescent microscopy, and fingerprinting, very distinct technological domains. In the idea sub-regime, designers/inventors source existing ideas (IOI) using analogical transfer and combine them probabilistically to create new ideas (IOI). Once new IOI are successfully created through probabilistic combination, they become part of the IOI pool, thus enlarging the number of ideas (IOI) in the pool for combination. It is important to clarify that model considers combinations at the ideas level rather combination of components, with the former being fundamental and allowing combination of ideas from different fields using analogical transfer.

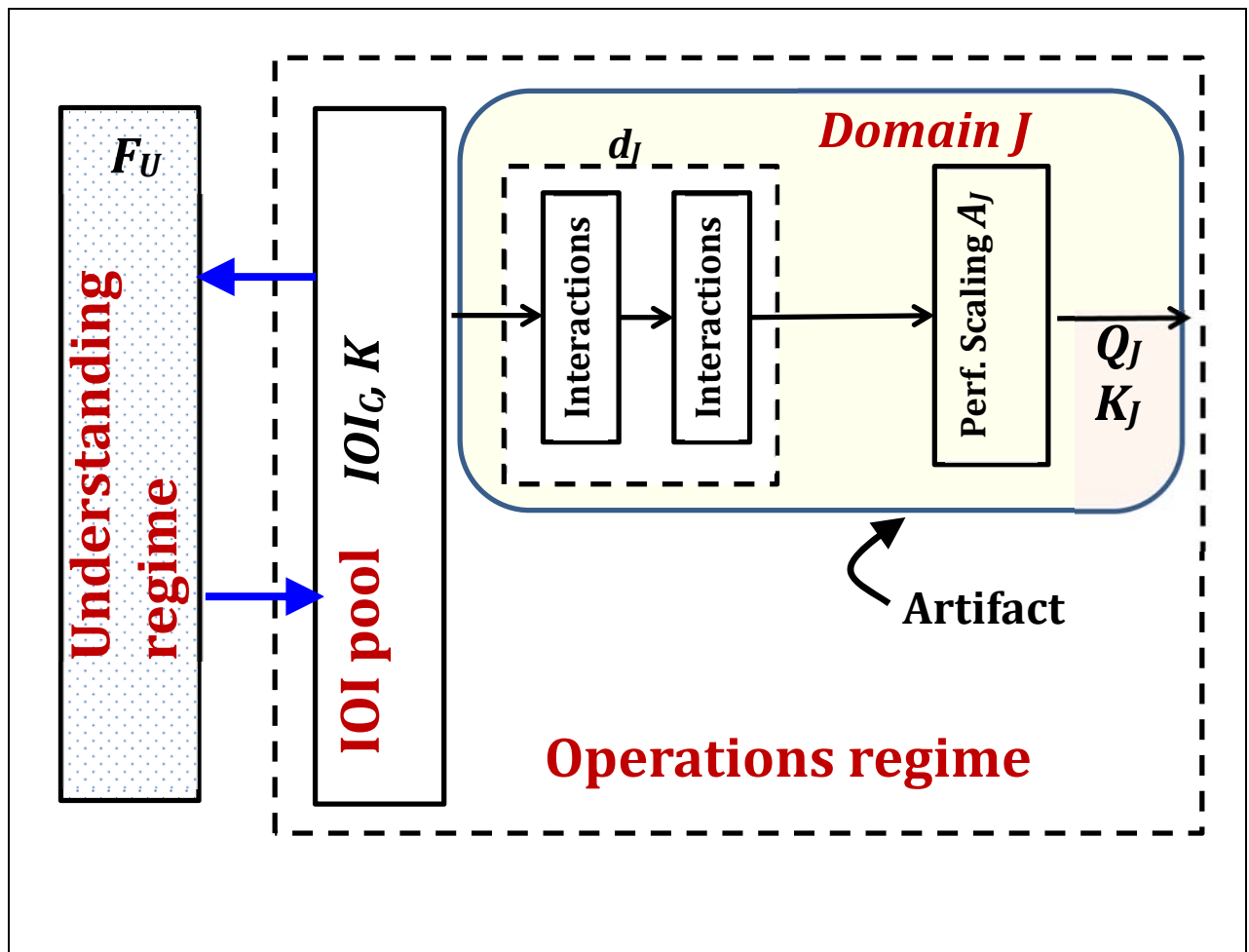

**Fig. 2: Model of exchange between Understanding and Operations regimes and modulation of IOI assimilation by interaction ($d_J$) and scaling ($A_J$) parameters of domain J.**

| Example of unit of understanding (UOU) | Example of incremental operating idea (IOI) |
|---|---|
| 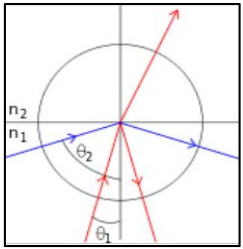 Principle of total internal reflection | 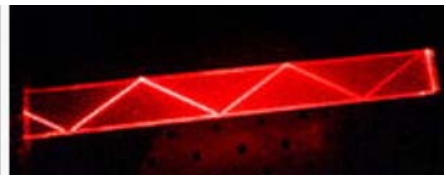 Total internal reflection between parallel surfaces enclosing dense medium: mechanism to make light travel longitudinally (fiber optics) |

**Fig. 3 Examples of unit of understanding (UOU) and incremental operating idea (IOI)**

We model growth in the explanatory reach of the Understanding regime by simulating a similar combinatorial analogical transfer process. The Understanding regime is conceptualized to consist of units of understanding (UOU). The units of understanding (UOU) from different fields within the understanding regime participate to create a new unit of understanding (UOU) that potentially (probabilistically) has a greater level of explanatory and predictive power. Following the treatment in Axtell et al. (2013), we model the explanatory and predictive power of a field of Understanding as a fitness parameter, $f_i$. If the new UOU has a greater fitness value, it replaces the UOU with the smallest fitness value. Since our primary focus is on performance - the output of the Operations regime, we simulate the Understanding regime only at this higher abstraction level.

Although both regimes – Understanding and Operations – evolve independently, they cannot do so indefinitely. We model the de Solla Price and Gribbin insights by having each regime act as a "barrier-breaker" for the other regime. When each regime hits a barrier, the other can eventually aid in breaking the barrier: infusion of understanding enables creation of important IOI in the Operations regime; and infusion of new operational tools enable new discoveries in the Understanding regime.

The performances of the artifacts in technological domains are improved by a series of designs/inventions (IOI) over time. IOI enable designers to change specific components in the domain artifact leading to a potential improvement. Following McNerney et al.'s treatment, the IOI in question is assimilated only if the performance of the artifact overall improves.

Another, and final, factor that we model is scaling, a property inherent in the physics of the design of the artifact.[5,6] The successfully assimilated IOI, which we refer to as $IOI_S$, effect improvement of the domain artifact by enabling favorable change of a relevant design parameter. The design parameter is increased or decreased such that it leads to improved performance[7]. Scaling refers to how change in a design parameter relates to relative change in the performance of an artifact. The formulation we use in the model is that relative performance change is related to design parameters raised to some power, in other words scaled. As covered in section 2.1, this is the most widely used functional relationship with decent empirical support and theoretical justification in some cases (Barenblatt 1996).

---

[5] Recall that the performance we consider in this paper is intensive, e.g., energy density, w/cm3.

[6] In relations to artifacts such as software, physics refers to the mathematics behind the software.

[7] Taguchi (1992) noted that some phenomena tend to work better when carried out at a smaller scale ("smaller is better"), while other are better at larger scale ("larger is better"). Integrated circuits, for example, perform better as dimensions are reduced, since smaller dimensions lead to shorter delays, and higher density of transistors, both of which contribute towards improved computation per volume or cost.

## 3.2 Mathematical summary

A performance (intensive) metric of a domain, labeled $Q_J$, is a function of a set of design parameters ($s_1, s_2, s_3$) of a domain artifact and time but for simplicity here we consider only a single parameter ($s$). The design parameter is changed by IOIs (successfully assimilated IOI into domain artifacts), which in turn are assimilated from $IOI_C$ (number of accumulated operating ideas in the IOI pool shown in Figure 2). $IOI_C$ is a function of time. Equations describing these nested variables in logarithmic form are:

$$\ln Q_J = f_1(\ln s); \ln s = f_2(\ln IOI_{SC}); \ln IOI_{SC} = f_3(\ln IOI_C); \ln IOI_C = f_4(t) \qquad (3)$$

Assuming that the functions are continuous and all dependence is through the named variables, the chain rule is applied and yields

$$d \ln Q_J / dt = d \ln Q_J / d \ln s \cdot d \ln s / d \ln IOI_{SC} \cdot d \ln IOI_{SC} / d \ln IOI_C \cdot d \ln IOI_C / dt \qquad (4)$$

The first term on the right hand side represents relative impact of design variable change on performance change, which will be shown in section 4.5 to be equal to the scaling parameter ($A_J$) when $Q_J$ follows a power law in $s$: $d \ln Q_J / d \ln s = A_J$. The second term is the 'smaller-is-better/larger-is-better' factor, and captures the notion whether a design variable has to be increased or decreased in order to improve performance. We capture this dependence using an abstraction and equate $d \ln s / d \ln IOI_{SC} = +/-1$.

Thus, equation (4) becomes

$$d \ln Q_J / dt = A_J \cdot (\pm 1) \cdot d \ln IOI_{SC} / d \ln IOI_C \cdot d \ln IOI_C / dt \qquad (5)$$

The third term on the right of equation (5) represents 'difficulty of implementing ideas' in specific domains, and thus relates the domain specific successful $IOI_{SC}$ to the $IOI_C$ in the pool: we will show in section 4.4 - following McNerney et al. - that $d \ln IOI_{SC} / d \ln IOI_C = 1/d_J$, where $d_J$ is the interaction parameter introduced by McNerney et al. Finally, the fourth term represents the rate of idea production. $K = d \ln IOI_C / dt$ is arrived at by a simulation of combinatorial analogical transfer which is presented in the first (following) section of the results.

# 4. Results

## 4.1 Overall IOI simulation

As noted in section 3.1, we model the IOI as resulting from *combining* knowledge from prior IOI by probabilistic analogical transfer. Fig 4a schematically represents combination of IOI, in which specific IOI a and b combine to create IOI d with a probability, $P_{IOI}$. If this combination attempt succeeds, the newly created IOI d then is added to the pool of IOI (Fig 4b). In subsequent time steps, IOI d can attempt to combine with another specific IOI in the pool, such as IOI c, to probabilistically create a more advanced IOI e. As combination advances, the cumulative number of individual operating ideas, $IOI_C$ grows. We further make the distinction between derived IOI and basic IOI, which we label as $IOI_0$. $IOI_0$ are fundamental IOI, which first introduce a natural effect into an operational principle to achieve some purpose. The example (described in section 3.1) of a pair of close parallel surfaces (or a fiber) enclosing a dense medium and utilizing principle of total internal reflection to transmit a beam of light longitudinally can be viewed as an example of an $IOI_0$. In contrast, derived IOI, just as the term suggests, are obtained through combination of two $IOI_0$, or between an $IOI_0$ and a derived IOI or between two derived IOI. In this sense, IOI a, b, and c in the figure represent $IOI_0$ and IOI d and e, derived IOI.

In one run of the simulation, we start with the initial number of basic individual operating ideas, $IOI_0$. At each time step, the maximum number of combinations we allow to be created is equal to half the number of total IOI available. The intention is to allow *each* operating idea to combine with another operating idea *once* per time step on average. Figure 5 shows results from a simulation run starting with 10 basic IOI and a probability of combination, $P_{IOI}$, equal to 0.25. Figures 5a and 5b with time steps on the X-axis and the cumulative number of operating ideas, $IOI_C$ on the Y-axis show that the cumulative number of operating ideas, $IOI_C$, grows exponentially with time at an improvement rate (K) of 0.116.

For this simplified case, the rate of growth of IOI, K, can be mathematically shown to be equal to $\ln(1 + P_{IOI}/2)$, =0.118 which can be easily derived as follows:

*At in a time step t, number of IOI newly created = $P_{IOI} \cdot IOI_C(t)/2$* (6)

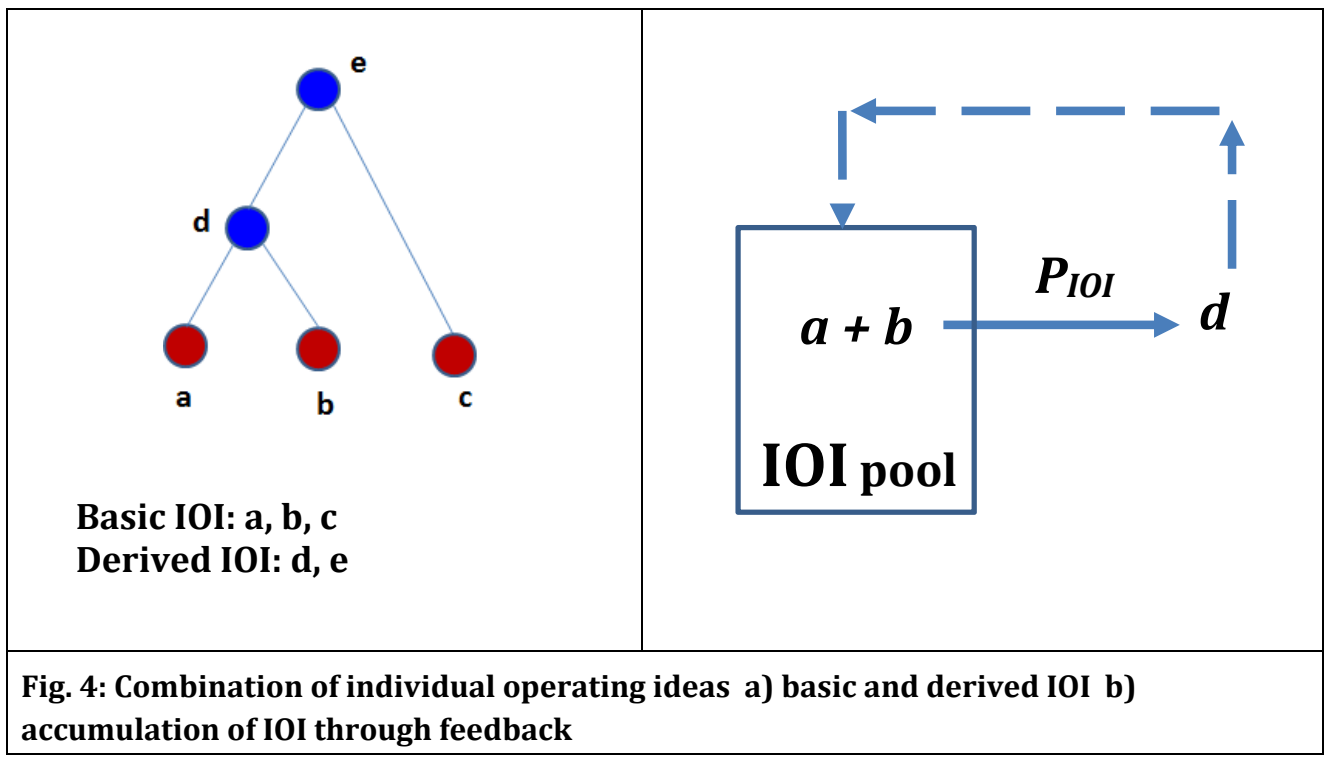

**Basic IOI: a, b, c**
**Derived IOI: d, e**

**Fig. 4: Combination of individual operating ideas  a) basic and derived IOI  b) accumulation of IOI through feedback**

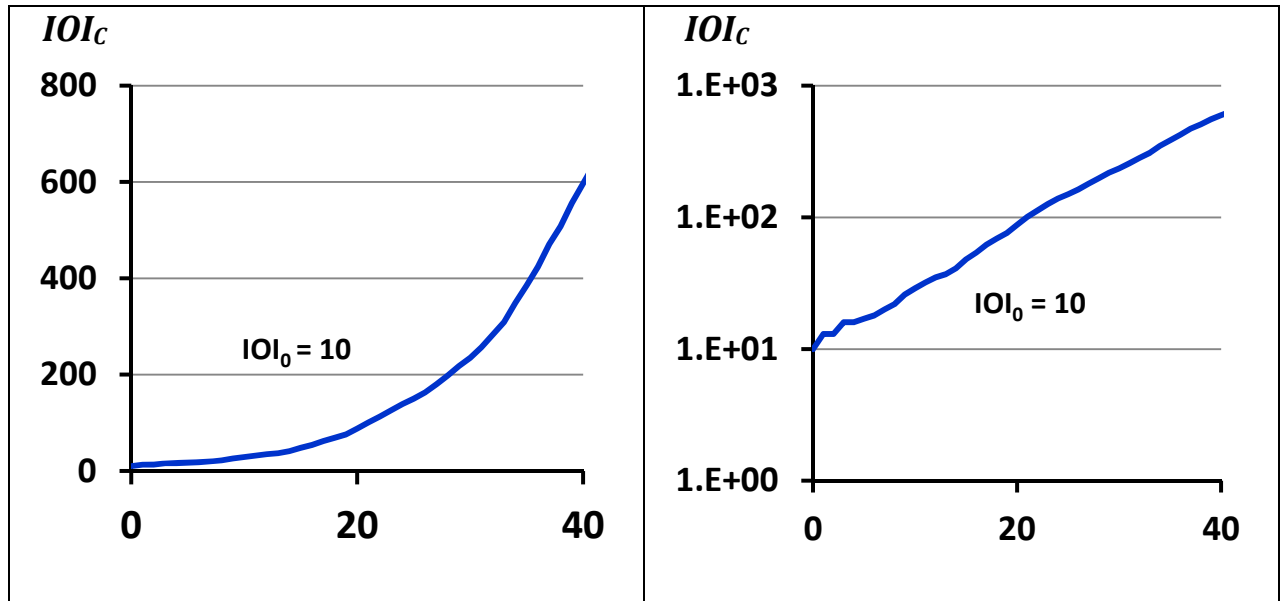

**Fig. 5: Growth of IOI$_C$ over time: initial IOI$_0$ = 10, probability of combination, $P_{IOI}$ = 0.25: (a) linear Y-axis (b) logarithmic Y-axis.**

$IOI_C(t+1) = IOI_C(t) + P_{IOI} \cdot IOI_C(t)/2 = IOI_C(t) \cdot (1 + P_{IOI}/2)$ (7)

Ratio of IOI$_C$ between consecutive time steps, $r = IOI_C(t+1)/IOI_C(t) = (1 + P_{IOI}/2)$ (8)

Then, in general, $IOI_c(t)$ can be written in terms of an initial IOI$_0$ and ratio, $r$ and time step, $t$; the expression can be stated in an exponential form.

$IOI_C(t) = IOI_0 \ r^t = IOI_0 \exp\{lnr \cdot t\} = IOI_0 \cdot \exp\{ln(1 + P_{IOI}/2) \cdot t\} = IOI_0 \cdot \exp\{k \cdot t\}$ (9)

Where, the rate of growth of IOI$_C$(t),

$K = ln(1 + P_{IOI}/2)$ (10)

For very small values of $P_{IOI}$,

$K \approx P_{IOI}/2$ (11)

The simulation results to this point assume that indefinitely large numbers of operating ideas, IOI, can be created out of few basic IOI. This is because the model assumes that the same operating ideas can be repeatedly used to create new IOI without limit. (For example, recombining (a,b) with a, then with b would give new operating IOI (((a,b),a),b) and eventually an arbitrarily large number of a, b pairs. Indefinite multiple uses of the same basic idea to create innumerable IOI does not appear to be realistic. In order to better reflect this intuition, we introduce a constraint that any derived IOI can utilize an IOI$_0$ only once. The constraint operationalizes the notion that counting repetitious use of basic IOI as new designs that potentially improve performance is unrealistic. According to this constraint, derived IOI ((a,b),c) in Figure 4 would be allowed, but not ((a,b),b). Employing this constraint, the simulation yields the results in Fig. 6a, a semi-log graph, showing the cumulative number of IOI initially growing exponentially with time. However, later on the curve bends over and hits a limit, demonstrating that all combination possibilities have been used up, and the pool of operating ideas stagnates which is also shown on the linear plot (Figure 6b) resembling a well-known "S curve".

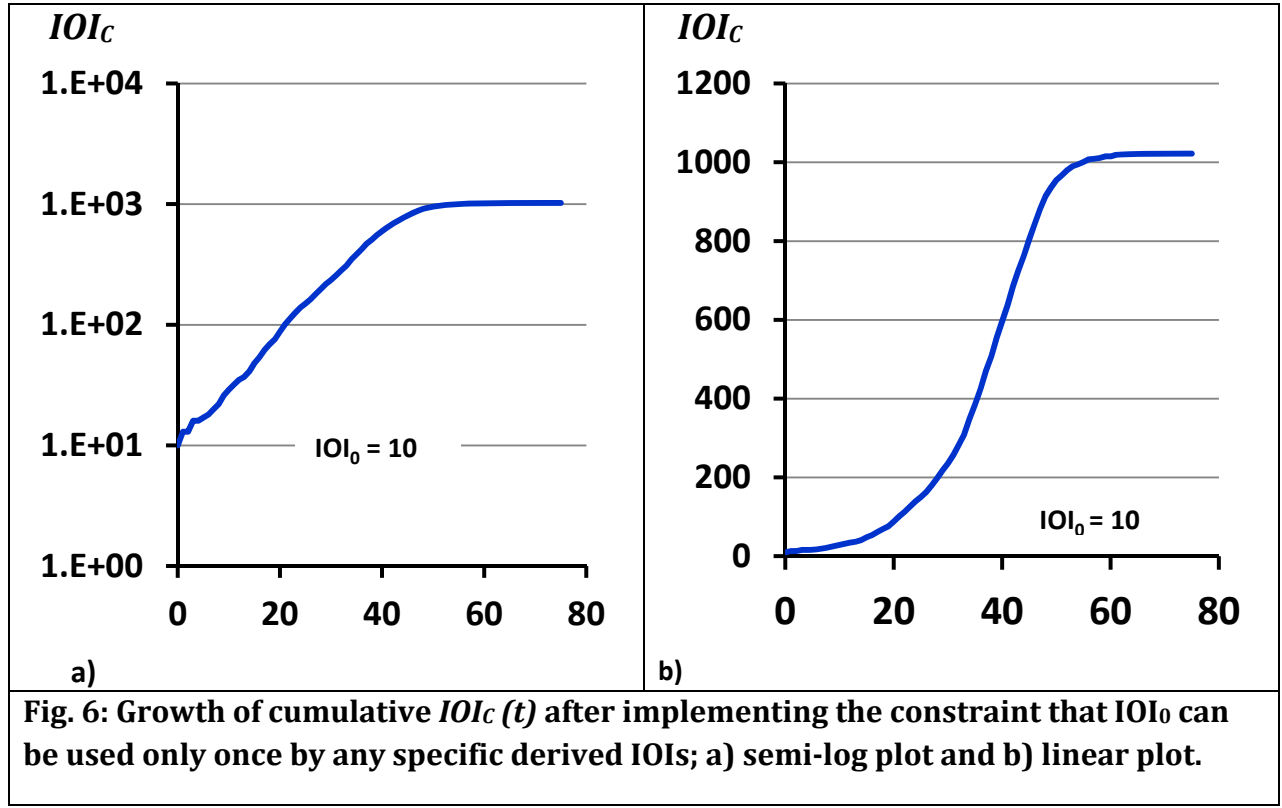

**Fig. 6: Growth of cumulative $IOI_C$ (t) after implementing the constraint that $IOI_0$ can be used only once by any specific derived IOIs; a) semi-log plot and b) linear plot.**

The maximum number of combination possibilities, which is a function of $IOI_0$ in the pool, defines the limit. This limit, or maximum number of combination possibilities, is given by a simple combinatorics equation (Cameron 1995):

$$IOI_{max} = 2^{IOI_0} - 1 \qquad (12)$$

Equation 12 entails that the limit increases rapidly as $IOI_0$ increases, due to its geometric dependence on $IOI_0$. For example, for $IOI_0$ equal to 5, 10, 15, and 20 the corresponding limits are 31, 1023 (Figure 6), 32767, and 1,048575 combination possibilities.

A natural question that arises from this result is what might determine the $IOI_0$ over time? We postulate a role for Understanding in this regard and we first briefly look at how Understanding evolves over time.

## 4.2 Combinatoric simulations for Understanding regime

Just like the Operations regime, we model the Understanding regime to also grow through a probabilistic analogical transfer process, in which units of understanding combine to create new units of understanding. In this model, we envision that the Understanding regime is composed of many fields, with each field having an explanatory reach. Using a treatment similar to the one used by Axtell et al. (2013), the explanatory reach of a field may be viewed as a fitness value of the theoretical understanding of that field, which we denote with $f_i$. Following Axtell et al., when units from two fields with fitness values, $f_1$ and $f_2$, combine, the fitness of the resulting unit is randomly chosen from a triangular distribution with the base or X-axis denoting the fitness values ranging from 0 to $f_1 + f_2$, and the apex representing the maximum value of the probability distribution function, given by $2/(f_1 + f_2)$. See Fig 7a. If the resulting fitness of the new understanding unit is higher than the fitness of either of the two combining units, the new understanding unit replaces the unit whose fitness is the smallest among the three. We assume the cumulative fitness of the Understanding regime ($F_U$) as a whole to be equal to the sum of the individual fitness value of each field.

Our simulation assumes 10 fields with starting fitness values ranging from 0 to 1, which are randomly assigned. Consequently, the average cumulative fitness ($F_U$) value is initially 5. As the simulation proceeds, fitness values of the 10 fields grow independently, and as a result, the cumulative fitness of the Understanding regime grows. Fig.7b shows results from a simulation run exhibiting roughly exponential growth of cumulative fitness over time. Thus, a simple model for growth of the Understanding regime is also exponential. However, as with the Operations regime, unlimited growth by simple combination of scientific theories is not realistic.

The Understanding regime also cannot progress by simple combination of existing understanding but instead experiences a limit that we envision as depending upon availability of operational (technological) tools available for testing scientific hypotheses and for discovering new effects. We express this dependence through an equation which expresses the maximum cumulative fitness at any time, $maxF_U(t)$, as simply proportional to the IOI existing at that time:

$$maxF_U(t) = Z_F \cdot IOI_C(t) \qquad (13)$$

Where $IOI_C$ thus represents an approximation for the effectiveness of available operational tools, and $Z_F$ is a constant of proportionality. This equation captures the concept first suggested by Price that the extent (or scope) of explanatory reach of the Understanding regime is dependent upon what experimental tools are available for scientists and researchers. It also recognizes in the terms of our model that these tools are essentially operational artifacts.

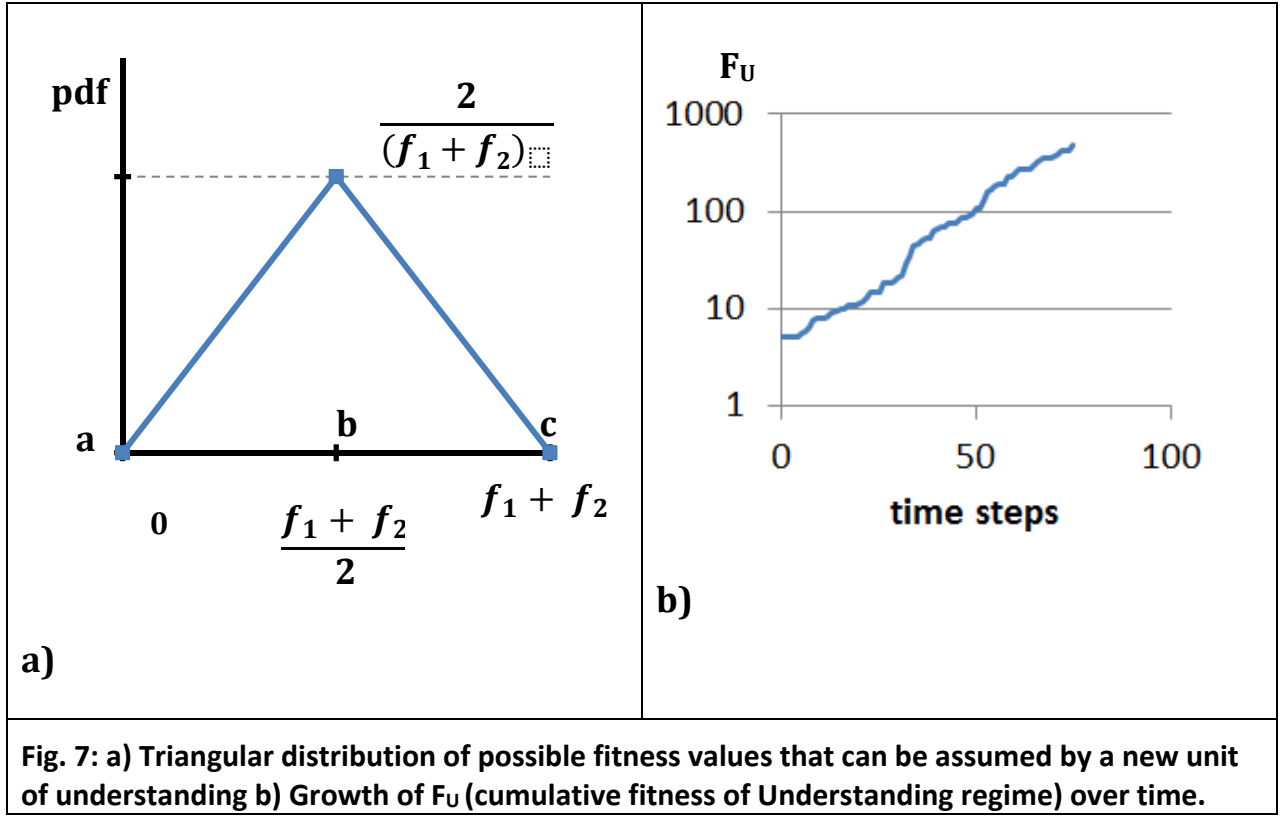

**Fig. 7: a)** Triangular distribution of possible fitness values that can be assumed by a new unit of understanding **b)** Growth of $F_U$ (cumulative fitness of Understanding regime) over time.

### 4.3 Exchanges between Understanding and Operations regimes

As discussed in section 3.1, prior qualitative work indicates that the interaction of Understanding and Operations is probably best modeled by assuming mutual beneficial interaction. In our model, we capture this enabling exchange from the Understanding to the Operations regime using a simple mathematical criterion:

$F_U(t)/F_U(t\_prev) \geq cutoff\_ratio\ (R)$ (14)

Where, $F_U(t)$ and $F_U(t\_prev)$ represent cumulative fitness values at time step t and the most recent time step, $t\_prev$, at which a $IOI_0$ had been introduced.

This criterion states that when cumulative fitness of the Understanding regime grows by some multiple ($R$) from the time when the last $IOI_0$ was invented, understanding has improved enough to generate a new $IOI_0$, which becomes available for combinations with all existing IOI. The threshold ratio, $R$, determines the frequency at which $IOI_0$ are created.

We now show results from a simulation including the exchange and limits on $IOI_0$. In the simulation, we study how synergistic exchange from Understanding influences the rate of growth of IOI in the Operations regime, including escape from stagnation. We focus particularly on two variables, namely, the initial number of $IOI_0$ in the Operations regime and the threshold ratio $R$ for creation of new $IOI_0$. Other pertinent variables are the probability of combination, $P_{IOI}$, the number of attempts per time step and the number of time steps per year and are not varied in this set of results.

For this simulation study, Table (1) presents the parameter values for $IOI_0$ (column 3) and the threshold ratios of cumulative fitness (column 4) that are used. As an example, 5B3R starts with $IOI_0$ of 5 and a new $IOI_0$ is created when cumulative fitness grows by a factor of 3. Both the initial number of $IOI_0$ and the threshold ratios of cumulative fitness are set at 3 different values, giving a total set of 9 parameter combinations. For all 9 runs, the probability for combination is kept constant at 0.25, and we assume one attempt per yearly time step.

Table 1: Simulation study: Parameter values of $IOI_0$ and $R$ (threshold ratios of cumulative fitness of Understanding) for the study. Results: $K$ is the slope fitting the simulation results to an exponential with $R^2$ for the fit (also shown). Other parameters, such as probability of combination, $P_{IOI}$ = 0.25, are kept constant.

|   | Simulation Run | Initial $IOI_0$ | Threshold ratio $R$ | Simulation avg. $K$ (± 2 std dev)[8] | $R^2$ | $K = ln(1+ P_{IOI}/2)$ |
|---|---|---|---|---|---|---|
| 1 | 5B1.5R  | 5  | 1.5 | 0.123 (± 0.011) | 0.998 | 0.118 |
| 2 | 5B3R    | 5  | 3.0 | 0.055 (± 0.019) | 0.959 | 0.118 |
| 3 | 5B5R    | 5  | 5.0 | 0.039 (± 0.007) | 0.943 | 0.118 |
| 4 | 10B1.5R | 10 | 1.5 | 0.122 (± 0.011) | 0.997 | 0.118 |
| 5 | 10B3R   | 10 | 3.0 | 0.115 (± 0.007) | 0.998 | 0.118 |
| 6 | 10B5R   | 10 | 5.0 | 0.117 (± 0.007) | 0.983 | 0.118 |
| 7 | 20B1.5R | 20 | 1.5 | 0.116 (± 0.007) | 0.998 | 0.118 |
| 8 | 20B3R   | 20 | 3.0 | 0.116 (± 0.009) | 0.998 | 0.118 |
| 9 | 20B5R   | 20 | 5.0 | 0.119 (± 0.016) | 0.998 | 0.118 |

---

[8] The standard deviation was estimated from seven repetitions for each simulation run.

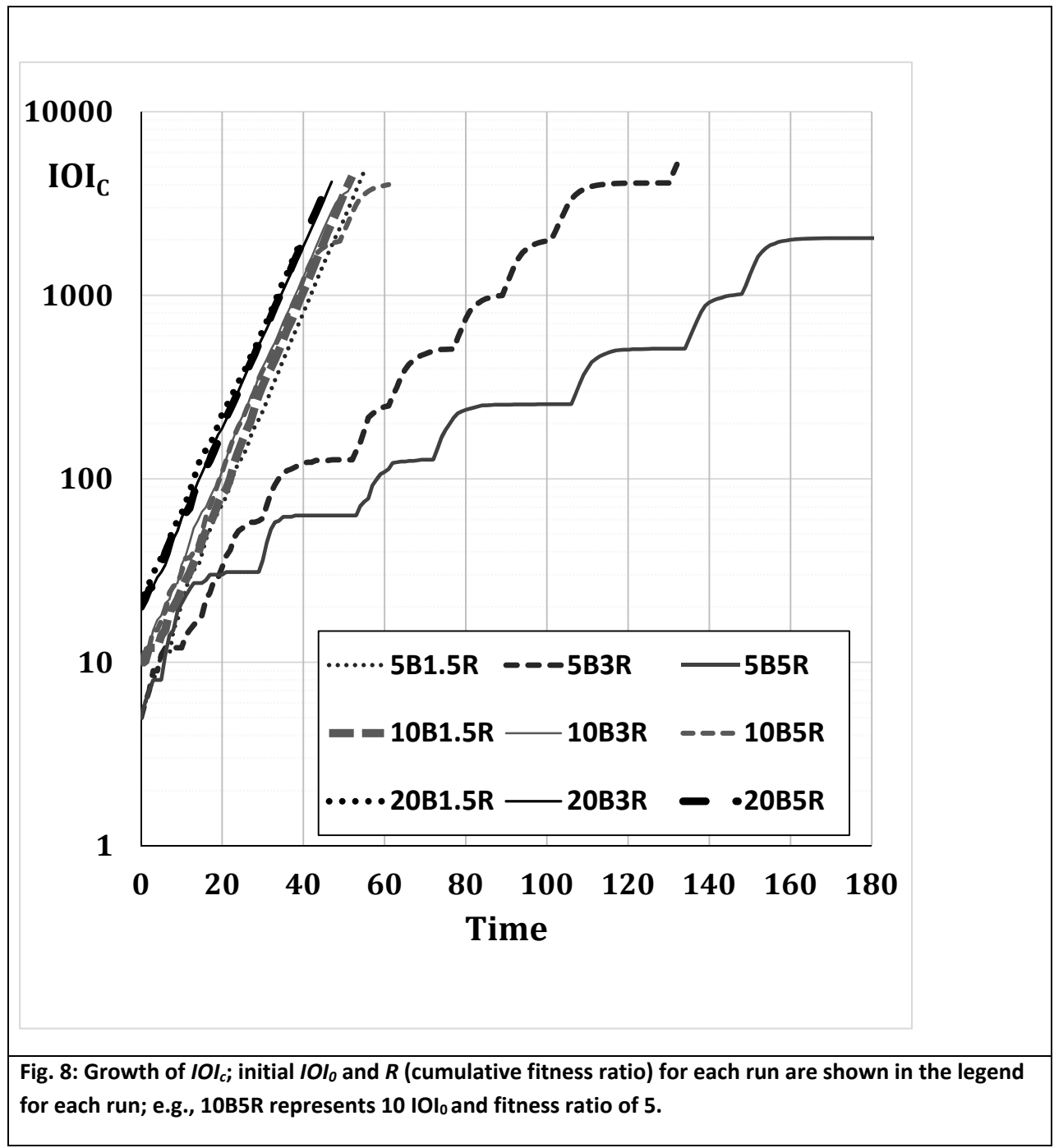

**Fig. 8:** Growth of *IOI$_c$*; initial *IOI$_0$* and *R* (cumulative fitness ratio) for each run are shown in the legend for each run; e.g., 10B5R represents 10 IOI$_0$ and fitness ratio of 5.

The simulation results in Fig. 8 shows the temporal growth of *IOI$_C$* in the Operations regime for the nine runs shown in Table 1. Runs 5B3R and 5B5R clearly stand out: they have a bumpy growth since they encounter periods of stagnation multiple times, as they evolve. Moreover, their effective rates of growth are meager, standing only at 0.055 and 0.04, which is much lower than 0.118, the rate given by Equation 10 {$ln (1+ P_{IOI}/2)$}. Columns 5, 6, and 7 list the *K*, *R$^2$*, and *K* calculated using $ln(1+ P_{IOI}/2)$ respectively. The small deviations from equation 10 found for the other 7 runs are within the 2-sigma estimated from multiple simulation repetitions for each run.

Both 5B3R and 5B5R start with low initial $IOI_0$ of 5 and have higher cumulative fitness threshold ratios ($R$) for infusion of new $IOI_0$. Low initial $IOI_0$ implies that the Operations regime has a low number of combinatorial possibilities of IOI to start with. Additionally, since new $IOI_0$ are not coming fast enough to push the frontier of combinatorial possibilities of IOI far enough, the Operations regime quickly exhausts the possibilities and again stagnates. Run5B5R stagnates for longer periods compared to 5B3R since it has a higher threshold ratio ($R$) for infusion of a new $IOI_0$ and thus slower progress. The Operations regime cannot escape the stagnation until another $IOI_0$ is created with infusion of new understanding. It is clear from the curves that this pattern repeats itself time after time.

Other simulation runs, except run 10B5R grow exponentially and smoothly and their rates are consistent with the theoretical value calculated using $ln(1+ P_{IOI}/2)$, 0.1178. These curves have either high enough $IOI_0$ to start with or fast infusion of $IOI_0$, or both. Run 5B1.5R, for example, starts with a low number of $IOI_0$ but has fast infusion of $IOI_0$, since the threshold ratio $R$ is only 1.5. On the other hand, run 20B5R has slow infusion of $IOI_0$ (high R), but starts with high initial $IOI_0$.

These runs do not exhibit stagnation for two reasons. The first reason is that the frontier of combinatorial possibilities for some runs is very far from the number of realized IOI at a given time step. For example, run 20B5R has over a million possibilities when it starts with 20 $IOI_0$. The second reason is that the frontier of the combinatorial possibilities keeps on moving further away as $IOI_c$ increases. Run 5B1.5R, for example, starts with 5 $IOI_0$, and yet it never experiences stagnation due to fast infusion of $IOI_0$ (low $R$) that push the frontier of combinatorial possibilities. The growth of $IOI_C$ is also free of stagnation for runs (e.g., such as Run10B3R) with medium number of initial $IOI_0$ and medium rate of infusion of $IOI_0$ (medium R). This is true because both factors in combination ensure that frontier of combinatorial possibilities is far enough to start with, and the frontier continues to move rapidly enough with time.

Run 10B5R exhibits somewhat unusual behavior. Although it grows smoothly at the beginning for quite some time, it experiences stagnation later on. This is because the frontier of combinatorial possibilities is far enough away to sustain steady growth early on. Later, the Operations regime exhausts the combinatorial possibilities before new $IOI_0$ arrive. However, once a new $IOI_0$ arrives, it jumpstarts again but it briefly halts at each new limit demonstrating the value of frequent interchange between Understanding and Operations in this simulation[9].

---

[9] The simulations are based upon infusion of $IOI_0$ depending upon a ratio ($R$) of growth in cumulative understanding, but similar results are found with assuming a model of difference in $F_U$.

We have seen that a combinatorial process combined with synergistic exchange between Understanding and Operations leads to an exponentially growing pool of operating ideas, $IOI_C$. This growth is described by an exponential function:

$$IOI_C(t) = IOI_0(t_0) \exp\{K(t - t_0)\} \qquad (15A)$$

$$K = \frac{d\, lnIOI_c}{dt} \qquad (15B)$$

Where, $K$ = the effective rate of growth of $IOI_C$, $IOI_0(t_0)$ = the number of initial basic IOI, $t$ = time, $t_0$ = initial time.

Our overall model (Section 3, Figure 2) envisages that this exponentially growing pool of operating ideas, $IOI_C$, provides the source for the exponential growth of performance of technological domains. How does this exponential growth of $IOI_C$ result in performance improvement and what accounts for the variation in rates of performance improvement across technological domains?

## 4.4 Modeling interaction differences among domains

As explained in section 3, two factors potentially responsible for modulating the exponential growth of operating ideas as they are integrated into technological domains are the domain interactions and scaling of relevant design variables. We consider domain interactions first following the work of McNerney et al. (2011) who modeled how interactions in processes affect unit cost. We build on their mathematical treatment to analyze the effect of interactions between components upon integrating an IOI into an artifact in a domain, which in turn improves the artifact's performance. Figure 9a shows a simplified schematic of an artifact in a technological domain that has three components (1,2,3) with interaction being depicted by out-going arrows, representing influence, from a component to other components, including itself. The outgoing arrows are referred to as out-links. The number of out-links, *d*, from a component provides a measure of its interaction level, and has value of 1 or greater as McNerney et al. assume each component at least affects itself. For simplicity, Figure 9a shows each component with two out-links, to itself and to another component. We represent an instance of an attempt being made to improve the performance of component 2 by an IOI being inserted. Since component 2 interacts with itself and another component, the performance of the interacting component is also changed by the insertion but in a fashion described probabilistically. The performance improvement attempt is accepted, only if the performance of the artifact as a whole improves. If that does occur, we follow McNerney et al. and consider the interactions being successfully resolved to improve the performance.

For a simplified artifact with *d* number of out-links for each component (*d*=2 in Fig. 9a), McNerney et. al.'s treatment (2011) for unit cost results in the following relationship:

$dC/dm = - B \cdot C^{d+1}$ \qquad (16)

Where, $C$ = unit cost normalized with respect to initial cost[10], $m$ = number of attempts, $d$= number of out-links, $B$ = constant

This equation states that the level of interaction inherent in the domain artifact influences the rate of unit cost reduction. We adapt this equation for our analysis in the following manner. We interpret the number of attempts as *IOIc* since the number of IOI determines the attempts (at each attempt an IOI is being introduced into an artifact to make a design change). Secondly, cost reduction is inversely related to performance improvement, such as in a typical metric kWh/$.[11] With these extensions of McNerney et al. equation 16 can be re-written as:

$d(Q)/d\ IOI_c = B \cdot Q^{-(d-1)}$ (17)

Where, $Q$ = performance

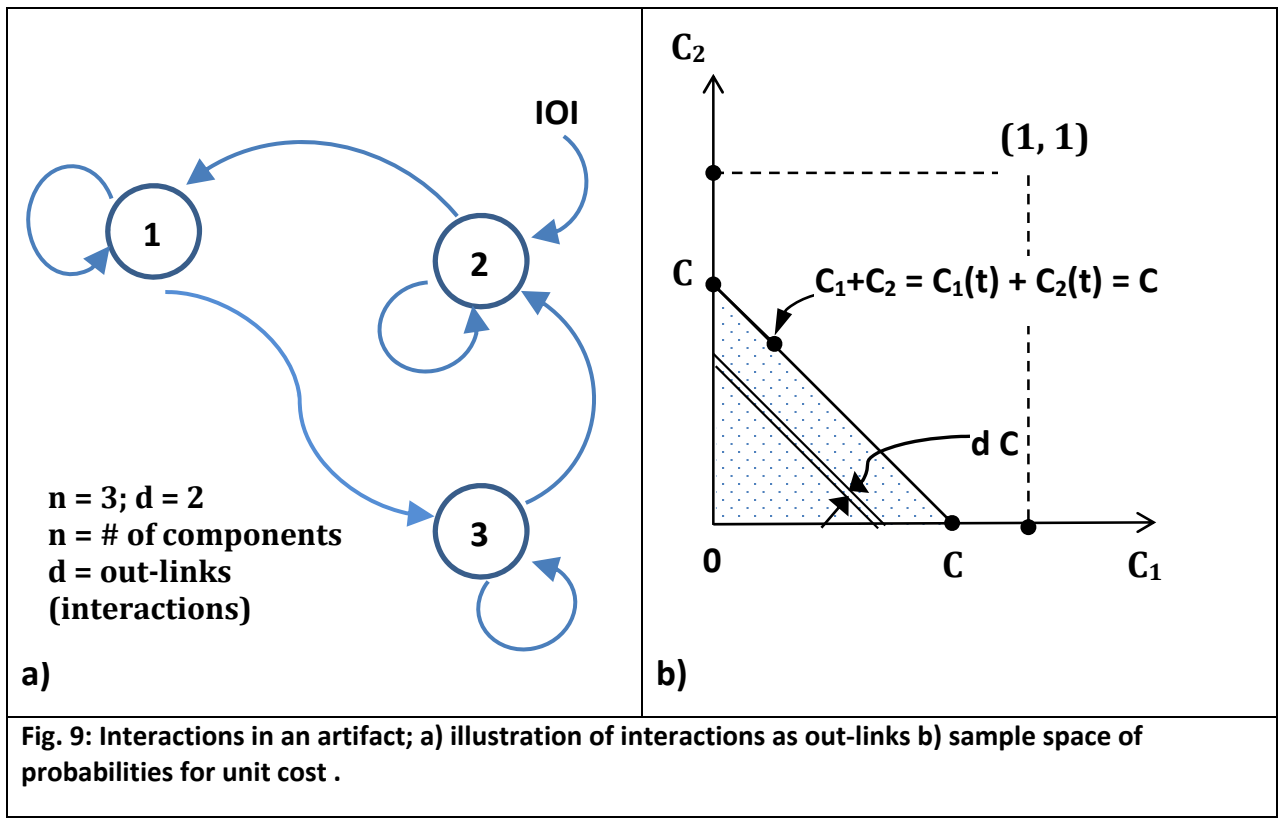

**Fig. 9: Interactions in an artifact; a) illustration of interactions as out-links b) sample space of probabilities for unit cost .**

Since as shown in Equations 4 and 5, successfully resolved operating ideas in a domain, $IOI_{SC}$, are the source for its performance improvement, we replace performance $Q$ of a domain with $IOI_{SC}$. An IOI is considered a successful attempt if the interaction resolution leads to net performance improvement of the artifact, and the count of successful IOI is denoted by $IOI_{SC}$. The modified equation shown below states that the interaction level, $d$, has a retarding effect on the growth of $IOI_{SC}$ in a domain.

---

[10] The normalized unit cost is 1 or less so increases in $d$ in equation 16 result in less improvement per attempt.
[11] The concept can be further generalized to include performance metrics which involve other resource constraints such as volume, mass, and time, instead of cost (e.g., kWh/m$^3$).

$d(IOI_{Sc})/dIOI_c = B \cdot IOI_{Sc}{}^{-(d-1)}$ (18)

We solve the differential equation by separating the variables ($IOI_{SC}$ on the left and $IOI_C$ on the right), and integrating both sides using dummy variables, and express $IOI_{SC}$ explicitly. The integration limits are: a) for the right side, 0 to $IOI_C$, b) for the left side, 1 to $IOI_{SC}$. The result is:

$$IOI_{SC} = (B \cdot d \cdot IOI_C + 1)^{1/d}$$ (19)

Since B and d are close to unity, and $IOI_c \gg 1$, we can ignore 1 in the brackets. Since our goal is to determine $\{d\ lnIOI_{SC}/d\ lnIOI_C\}$, we take the natural log of both sides and differentiate it with respect to $ln\ IOI_C$, resulting in the following expression which will be substituted into equation 5 in section 4.6:

$d\ lnIOIsc\ /\ d\ lnIOIc\ = 1/d_J$ (20)

### 4.5 Performance models - scaling of design variables

Our research question is concerned with intensive technological performance of domain artifacts. The intensive technological performance represents an innate performance characteristic of an artifact. We operationalize the notion of intensive performance by dividing desirable artifact outputs with resource constraints (e.g., mass, volume, time, cost). An intensive performance metric for batteries is energy density, $kWh/m^3$. We now consider three examples of relationships between intensive performance and design variables.

### 4.5.1 Selected examples

We first consider blast furnaces used in the manufacturing of steel as representative of reaction vessels of various kinds. Widely used performance attributes for a blast furnace are capacity and cost, where cost can be considered the resource constraint. So, an intensive performance metric can be defined as capacity (output per hour or day typically) per unit cost. The capacity of a reaction vessel is proportional to its volume while its cost is primarily proportional to surface area (Lipsey et al. 2005). The following dimensional analysis shows that following these simplistic assumptions, intensive performance of a reaction vessel is linearly proportional to size, s.

$Q_{RV}$ = *capacity/cost of reaction vessel* = $s^3/s^2 = s^1$ (21)

Gold (1974) has empirically shown that the cost of a blast furnace goes up by 60 percent when the capacity is doubled. Intensive performance $Q_{RV}$ using this empirical finding goes up by 1.25 (=2/1.6) when $s^3$ doubles, and thus $s$ goes up by 1.26 (=$2^{.333}$) closely agreeing with the simply derived equation 21.

A second example we consider is specific power output from internal combustion (and other heat) engines. Power output (kW) is proportional to volume occupied by the combustion chamber minus the heat loss from the engine, which in turn is proportional to the engine's surface area. The power, then, is:

$$power = A s^3 - B s^2 \; ; \; B/A < 1 \qquad (22)$$

Where *A* and *B* are constants for power generation and heat loss respectively.
$Q_{IC}$ = specific power $\alpha$ power/volume of engine; thus specific power is

$$= (A s^3 - B s^2)/s^3 = A - B/s \qquad (23)$$

Equation 23 indicates that, similar to reaction vessels, specific power output of IC engines increases with size so both are "larger is better" artifacts". For small values of *B/As*, specific power increases approximately linearly with *s*. For larger values of *s*, the increase is less than linear in *s*.

As a final example, we consider information technologies, whose performance improvement ranks amongst the highest. Several modern information technologies depend upon integrated circuit (IC) chips. Electronic computers have been improving performance by reducing the feature sizes of transistors in IC chips for microprocessors. The number of computations per second per unit volume, an intensive measure of performance, depends upon frequency and the number of transistors in a unit volume. Frequency is inversely proportional to the linear dimension of a feature, *s*, and the number of transistors per unit area is inversely proportional to area of the feature. Thus,

*Computation per sec per cc* $= 1/s \cdot 1/s^2 = s^{-3}$ \qquad (24)

The dimensional analysis indicates that computations per second increases rapidly for a decrease in a linear dimension of a feature. This is due to the cubic (or higher)[12] dependence of computations per second on feature size. The negative sign captures the fact that reduction of the design variable increases performance – smaller is better for this artifact.

### 4.5.2 Generalization of scaling of design variables
The three examples we have presented illustrate the notion that intensive performance improved by different degrees depending how the design variables are scaled. In the first two cases, a 10 percent increase in a design variable will improve performance by 10 percent or less. However, in the case of computations, for the same 10 percent change in design variable (feature size), the performance would improve by over 33 percent. This dependence is modeled as a power-law[13]:

---

[12] If the vertical dimension also decreases over time as the feature size decreases, a higher power-perhaps approaching 4 - would apply.
[13] The engine example demonstrates that this is an approximation in many cases.

$$Q_J = s^{A_J} \quad (25) \qquad \ln Q_J = A_J \ln s \quad (26)$$

$$d \ln Q_J / d \ln s = A_J \quad (27)$$

Where, $A_J$ is the scaling factor for domain $J$, $s$ is the design variable.

### 4.6 Bringing all elements together

We now bring the results for rate of $IOI_{SC}$ growth and influence of interaction and scaling together. For the reader's convenience, we reproduce equation 4 here, and substitute the results for the four factors:

$$d \ln Q_J/dt = d \ln Q_J/d \ln s \cdot d \ln s/d \ln IOI_{SC} \cdot d \ln IOI_{SC}/d \ln IOI_C \cdot d \ln IOI_C/dt \quad (4)$$

Substituting the results from equations 27, 20, and 15B for the first, third and fourth terms, ±1 for the second term, and then rearranging, we get:

$$K_J = \frac{d \ln Q_J}{dt} = (\mp 1) A_J \; \frac{1}{d_J} \; K \quad (28)$$

Equation 28 represents the overall model of the annual rate of improvement for domain $J$. According to this equation, $K_J$, the annual rate of improvement of domain $J$ depends upon $K$, the exponential rate at which the $IOI_C$ pool increases in size. $K$ is then modulated by domain specific parameters, $d_J$ (interaction) inversely and $A_J$ (scaling) proportionally to result in a domain specific rate of improvement $K_J$. The minus sign is converted into positive one by negative sign of $A_J$ (for those cases where smaller is better). One observation to note is that $A_J$ and $d_J$ are constants for a given domain, thus resulting in a time invariant rate (or a simple exponential) for a domain.

## 5. Discussion

The goal of this paper was to develop a mathematical model that utilizes mechanisms in the design/invention process to examine the nature of technological performance improvement trends. The exploration has utilized simulation to gain insight into a combinatorial process based upon analogical transfer and Understanding /Operations exchange and quantitatively modeled interactions and scaling. In this section, we first briefly review the consistencies of the model with empirical results (and what is known about technological change). All empirical results we are aware of are found to support the model. We then consider the as yet untested predictions from the model as well as the assumptions made in the model.

According to the model, the exponential nature of performance improvement for all technological domains arises in the idea realm of the operational knowledge regime, where new inventive ideas are created using combinatorial analogical transfer of existing ideas, which, in turn, become the building blocks for future inventive ideas. We emphasize that the combinations modeled are occurring at the idea level, although combinations can also take place between components. As noted in section 3.1, we make this distinction as the former is much more pervasive and allows combination of ideas from different fields; however, it is likely that some ideas cannot be combined and this is treated probabilistically since many combination attempts fail. The model demonstrates this incessant cumulative combinatorial aspect of knowledge in both the Understanding and the Operations regimes manifests as exponential trends. The combinatorial model is simple but it leads naturally to the exponential behavior with time that has only been obtained previously by Axtell et al. in a model that went beyond performance to diffusion over a set of agents. Since such exponential behavior with time is one of the most widely noted behaviors of technical performance (Moore 1965, Koh and Magee 2006, 2008, Nagy et al. 2013, Magee et. a. 2014), the combinatoric model enacting analogical transfer that was developed in the current paper is clearly supported by what is known empirically about performance trends with time.

The Operations and the Understanding regimes can improve independently in the model but not indefinitely. How long the Operations regime can improve depends in the model upon the size of the technological possibility space, which according to the model is dependent on the number of basic IOI, fundamental operational principles, existing. The Understanding regime can also experience stagnation, but this happens when the operational tools that scientists and researchers use for discovery and testing hypotheses are not adequate. The Operations regime comes to its rescue by providing these operational tools in form of empirical methods, tools and instruments (increased numbers of individual operating ideas), which greatly enhances the scientists ability to discover and test, and thus further push the limits of understanding in the manner suggested by Price (1983), Gribbin(2002) and in the following quote from Toynbee (1962).

> *Physical Science and Industrialism may be conceived as a pair of dancers both of whom know their steps and have an ear for the rhythm of the music. If the partner who has been leading chooses to change parts and to follow instead there is perhaps no reason to expect that he will dance less correctly than before.*

In this sense, the Operations regime and the Understanding regime are like two independent neighbors who interact for mutual benefit. In the model, their frequency of interaction however influences their effective rate of growth. Our model is a specific realization that achieves this mutual interaction that has previously been widely noted from deep qualitative research.

The results in Figure 8 are summarized as a surface plot in Figure 10. *K*, the effective rate of growth of $IOI_C$ was determined by the initial $IOI_0$, and the frequency of interaction ($\alpha$ $1/ln\ R$). The former determined the envelope of technological possibility space. When *$IOI_0$* are high, the effective rate of growth *K* is close to the theoretical combinatorial rate determined by Equation 10 { *$ln(1+ P_{IOI}/2)$* }, irrespective of whether there was frequent exchange. However, when the *$IOI_0$* are low, the limit is hit repeatedly, translating into
30

halting and a reduced effective rate of growth. The value of *K* in this case was determined by the frequency of enabling exchange from the Understanding regime, with higher frequency (low *R*) leading to higher effective rate. With sufficiently high frequency, even with low initial *IOI$_0$*, the effective rate *K* eventually approaches the theoretical rate.

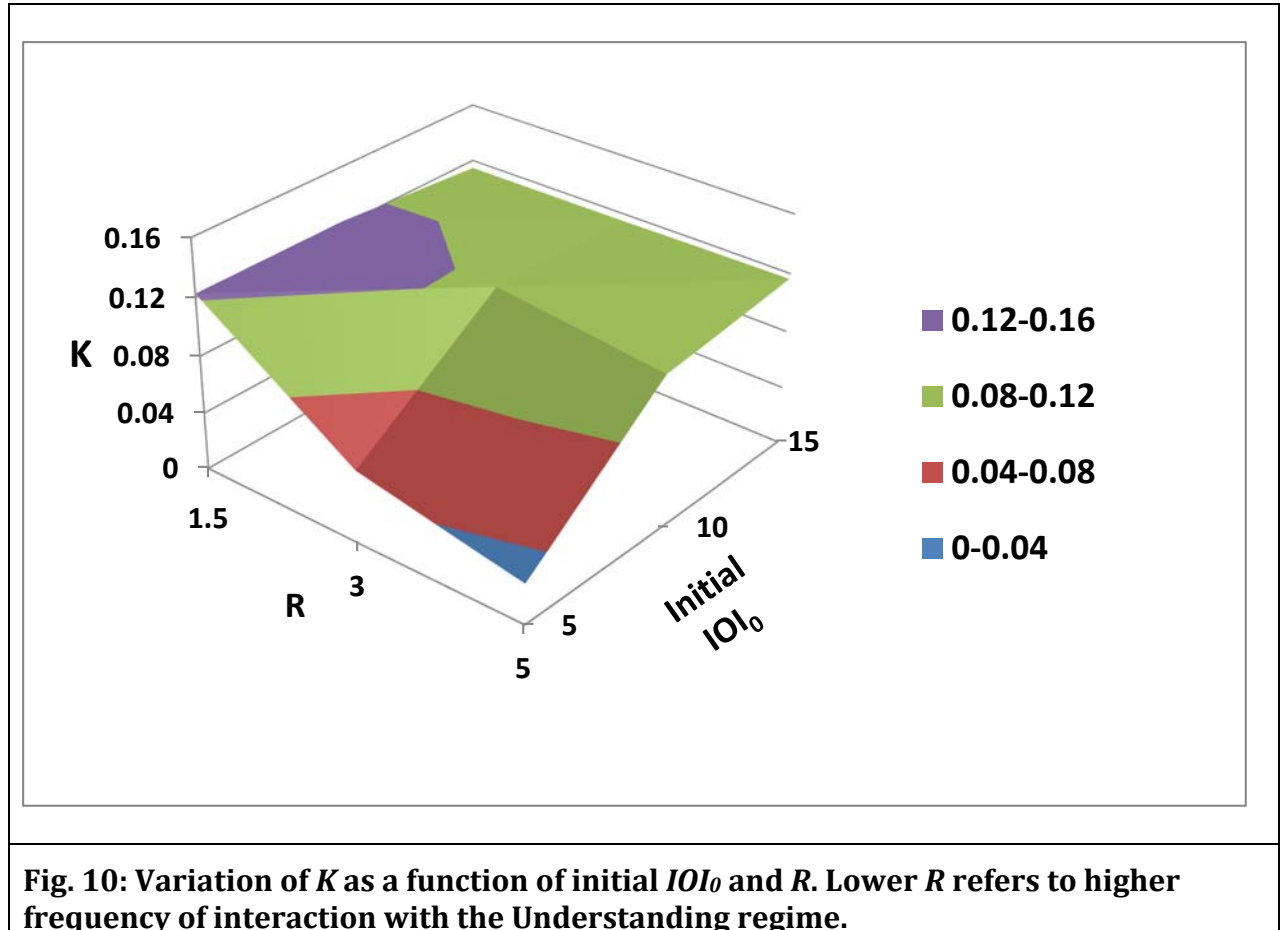

**Fig. 10: Variation of *K* as a function of initial *IOI$_0$* and *R*. Lower *R* refers to higher frequency of interaction with the Understanding regime.**

Detailed historical studies of technological change (Mokyr 2002) note centuries of slow, halting progress that eventually becomes much more rapid and sustained starting in the late 18th century in the UK. An interesting consistency of these observations with our model is seen since our model attributes the transition to sustained higher improvement rate to the combinatorial growth of individual ideas that are able to reinforce one another by the analogical transfer mechanism. That our model partially accomplishes this through the synergistic exchange between Understanding and Operations is also consistent with the detailed historical studies as interpreted by many observers (Schofield 1963, Musson 1972, Rosenberg and Birdzell 1986, Musson and Robinson 1989, Mokyr 2002, Lipsey et al. 2005).

The K$_J$ values found empirically vary by approximately a factor of 22 (from 0.03 to 0.65 according to Magee et al. (2014). Equation 28 states that annual improvement rate for a domain is determined by the product of *K* times the scaling parameter, $A_J$, and the reciprocal of the interaction parameter, $d_J$. According to this result, the last two parameters produce the variation of improvement rates across domains. During the embodiment process, interactions prevalent in the domain artifacts influence how many inventive ideas can be absorbed. The percent increase in successfully absorbed ideas by a domain artifact is inversely proportional to the average interaction parameter of the domain $d_J$. By definition, the minimum value of d is 1 and the maximum might be higher but a value of 6 appears reasonable. The other factor that is predicted to differentiate domains is performance scaling. Inventive ideas affect artifact performance by modifying the design parameters in domain artifacts. The model indicates that the relative improvement of performance for a given number of absorbed new operating ideas is governed by the scaling parameter $A_J$. The examples presented in section 4.5 illustrated that the value of $A_J$ can vary across domains. In particular, for the IC domain (where smaller is better), $A_J$ is apparently 3 to 4 times larger than for typical larger-is-better domains such as combustion engines. Thus, the range of $K_J$ empirically observed is potentially explainable by changes in $d_J$ and $A_J$, but much more empirical work is needed to fully support these quantitative implications of Equation 28 as will be discussed further below.

The empirical findings of Benson and Magee (2015a) also support the model. In particular, they found no correlation of rates in domains with effort in a domain (measured by number of patents or patenting rate) or with the amount of outside knowledge used by a domain (this is very large for all domains). They interpreted their findings by a "rising sea metaphor" that represents all inventions and scientific output being *equally* available to all domains but that fundamentals in the domains determine the rate of performance improvement. Overall effort in Understanding (science) and invention increase the rates in all domains but the differences among rates of improvement are due to differences in fundamental characteristics among the domains. The model in this paper identifies interactions and scaling as two such fundamentals and equation 28 is specific about the variation expected due to these two fundamental characteristics.

Thus, our model is supported by what is known empirically including exponential dependence of performance on time; slow, halting progress in the early stages of technological development; a role for science in enabling technological performance improvement; the range of variation in performance improvement across domains; and the importance of domain fundamentals to variation in performance. However, to what extent does it achieve the ideal level of understanding mentioned in section 2 when discussing the related Benson and Magee research? It is - as desired - based upon what is known about the design/inventive process and does not rely upon characteristics only determined by observation of output in a domain. Moreover, it provides explanations of existing empirical results not made by prior models. However, does it make any new predictions; do its assumptions appear reasonable; and what new avenues of design research, if any, does it open up for further exploration? We consider these issues in the remainder of the discussion.

There are three new predictions made by the model as instantiated in Equation 28. These are: 1) that the noise in estimating $K_J$ should vary with $K_J$ linearly rather than for example be independent of $K_J$; 2) that performance improvement comparisons across domains vary as $1/d_J$ where d is the interaction parameter; and 3) that performance improvement across domains vary as $A_J$. The first prediction follows from the fact that the model ascribes all variation in the process to the probabilistic analogical transfer process that creates IOI and thus any noise generated in the process is amplified by the same factors that determine $K_J$ (namely $1/d_J$ and $A_J$.). Very recent work appears to confirm the first prediction. In a careful study of the observed noise in a wide variety of domains, Farmer and Lafond have find that the variation in $K_J$ is proportional to $K_J$ offering empirical support to the form of Equation 28. This is potentially an important confirmation of a prediction of the model but the careful work by Farmer and Lafond has potential data limitations (detailed in their paper) and further work of this kind is highly desirable. Prediction 2 is that component interactions ($d_J$), which characterize the domains, influence improvement rate by modulating the implementation of IOI in the domain artifacts. This prediction can be tested by study of the performance improvement rates over a variety of domains where an independent assessment of $d_J$ is made. The authors have performed such a test using patent data (Basnet and Magee 2016) and the results, which demonstrate positive correlation between improvement rates *($K_J$)* and interaction parameter *($d_J$)*, offer support for the analysis of McNerney et al. that we use in our model. Prediction 3 is that relative improvement among domains varies proportionally to the scaling parameter for the domain design parameters, a consequence of performance following a power law with the design parameters. If scaling laws were found (or derived) for a variety of domains whose rate of progress is known, prediction 3 can also be tested. In this paper, we showed that the factor A is at least 3 times larger for Integrated circuits than for combustion engines. While this provides preliminary support for the model since Integrated circuits improve about 7 times faster than combustion engines (Magee et al, 2014), two points do not achieve a rigorous test. One would need to have reliable scaling factors for at least 10 domains with varying $K_J$ to determine whether this part of the model is empirically supported.

A fundamental aspect of the overall model is that it differentiates between the idea/knowledge and artifact aspects of design and invention. Such decomposition is an essential step in arriving at our key result (equation 28 through equation 5). It is not clear that this assumption is testable so it must remain an unverified assumption or definition but we do note that it appears to accord with reality in that inventors/designers spend significant amount of time working with ideas and representations of artifacts, for example in the form of sketches and drawings, well before they build artifacts. Others have noted the higher leverage of analogical transfer between ideas as opposed to designed artifacts (Weisberg 2006).

A potentially important and non-obvious assumption made in the model is that inventive effort increases as the cumulative number of individual operating ideas - $IOI_C$ - increases. This assumption is introduced when we assume that *every* existing IOI undergoes a combination attempt in *each* time step. As $IOI_C$ increases, this means that more inventions are attempted in each successive time step. This assumption is critical to obtaining the exponential time dependence for $IOI_C$ and thus for Q because the growth of $IOI_C$ would be choked off if inventive attempts did not increase over time. Although a rigorous test of this assumption is suggested for further work, we do note support for the assumption in the exponential growth of patents over time (Youn et al. 2014, Packalen and Bhattachayra, 2015)[14]. Approximate support is also given by the roughly exponential growth of R&D

---

[14] Both of these papers show more rapid exponential increases before 1870 and slower but still exponential increases over time from 1870 to the present in the number of US patents.

spending over time (NSF, 2014) and by the roughly exponential growth of graduate engineers globally[15] over time (NSF, 2014)

The model assumes a simple exchange between Understanding (largely science) and Operations (largely technology) as described by Equations 13 and 14. The details of this mechanism are not testable but in our opinion not critical because other formalisms (based upon differences rather than ratios and based upon count of units of understanding rather than our choice of explanatory reach) lead to results closely similar to those reported here. Therefore, this assumption remains unverified but is not critical to our conclusions. Similarly, the initial value of $IOI_0$ chosen in the simulation (and the exchange frequency with Understanding ($\alpha$ 1/ln $R$)) is essential to our finding of halting slow growth that can transition to sustained and more rapid growth. Although this finding is consistent with detailed observation as noted above and the initial number of useful ideas must be small, there is no independent means of assessing $IOI_0$. Moreover, we have made a number of assumptions in parameter values to construct a simple and operational simulation. The values for parameters in the simulation, such as $P_{IOI}$, number of time steps, number of scientific fields, $R$, fitness values are chosen to keep the computational cost reasonable, without sacrificing the essential aspects. Simulations show that results are robust to different combinations of parameter values with respect to exponential trends and variation in rates. Therefore, these choices and simplifications do not undercut the explanatory or predictive capabilities of the model but do limit the potential for non-calibrated calculation of, for example, the improvement rate for a domain since K is only approximately known.

To make the model tractable, we have made number of simplifying abstractions, introducing several other limitations to the model. Since the model is not agent-based, it does not distinguish between organizations nor between inventors. Since our goal is to explain the patterns at the domain level, we consider the domain as one entity. For this reason, variations among organizations or among inventors within a domain are not taken into account, and hence the model is not useful to understand organizational or individual inventor effectiveness in its current form and any systematic differences among inventor capability across domains is ignored. Second, once IOI are created by any inventor, the model assumes they are instantly available for combinatorial analogical transfer across the pool underlying all domains. Thus, the model does not take into account time delay that can result due to, for example, geography, secrecy and governmental regulations, and hence is not useful for studying such factors' influence in technological change. Third, the model assumes that 2 pre-existing ideas are sufficient (probabilistically) to create another idea whereas inventions also result from bringing more than 2 pre-existing ideas together. However, adding such complications to the model and simulation does not change the fundamental findings since the creation of new ideas would still increase as the number of pre-existing ideas increase as long as we still assume an increasing invention effort. Fourth, although conceptually the notion of fitness of scientific fields makes sense, how the fitness can be measured, and who measures it for a scientific field are contested, especially for rapidly growing fields.

---

[15] Other supporting evidence is also possible to see in the NSF material at http://www.nsf.gov/statistics/seind14/index.cfm/overview/c0s1.htm#s2

This analysis of the predictions points out that some key aspects of Equation 28 have the potential to be empirically tested and thus are clear future research activities suggested by the model. Among these future research activities, one important issue to discuss is the extensions possible to design research potentially opened up by the current work. The model in this paper explicitly considers design changes in succeeding artifacts in a series to be the central element in technological change over time. Thus, it adds to the few other papers (Baldwin and Clark 2006, Luo et al. 2014) that have connected these two large fields of research - technological change and design theory. This paper in particular connects design conceptually and quantitatively to changes in performance over time. Since there is significant data of this type (Moore, 1965, Girifalco 1991, Nordhaus 1996, Koh and Magee (2006, 2008) and Leinhard 2008), this paper points the way for further quantitative comparisons of models based upon design theory with data. Another line of research that this model suggests is more explicit consideration of interactions and scaling as part of design theories. The current model explores simple models for both of these that are capable of predicting differences in time dependence of performance in differing domains. Design of artifacts could conceptually be changed so that the potential for improvement with ongoing redesign is enhanced possibly through reduced interactions or more intensive scaling relationships. Thus, the current paper suggests the potential importance of further research on specific differences in design approaches with different scaling laws and with different level of interactions.

## 6. Concluding remarks

The model and simulations of the improvements in performance due to a series of inventions (new designs) over time presented in this work are based upon a simple version of analogical transfer as a combinatorial process among pre-existing operational/inventive ideas. The model is supported by a number of empirically known aspects of technological change including:
1. The transition from slow, hesitant technological change to more sustained technological progress as technological ideas accumulate;
2. A role for the emergence of the scientific process in stimulating the transition in point 1;
3. The exponential increase of performance with time (generalized Moore's Law) seen quite widely empirically;
4. That stochastic noise in the slopes of the log performance vs. time curves is proportional to the slope;
5. The level of effort in domains is not important in the rate of progress.

The model also indicates that:
6. The rate of performance increase in a technological domain is at least partly (and possibly largely) due to fundamental technical reasons (component interactions and scaling of design variables), rather than contextual reasons (such as investment in R&D, scientific and engineering talent, or organizational aspects).

Numerous modeling assumptions were made in developing the model but only some of these are critical to the conclusions just listed. Further specific research is suggested to move some critical assumptions into the testable category, and to consider interactions and scaling parameters in new design approaches. These are discussed in the paper particularly for the assumptions underlying point 6 above. The tests involve detailed studies of the interaction and scaling parameters in a variety of domains. All of this future research could support or lead to modification of point 6.

## Acknowledgement

The authors are grateful to the International Design Center of MIT and the Singapore University of Technology and Design (SUTD) for its generous support of this research. We would also like to thank Dr. James McNerney for helpful discussion about artifact interactions. We want to also acknowledge valuable input on an earlier version of this paper by Dr. James McNerney and Dr. Daniel. E. Whitney.

# Reference

Acemoglu, D. (2002). Directed Technical Change. *The Review of Economic Studies*, *69*(4), 781–809. http://doi.org/10.2307/1556722

Arrow, K. J. (1962). The economic implications of learning by doing'. *The Review of Economic Studies*, *29*(3).

Arthur, W. B. (2007). The structure of invention. *Research Policy*, *36*(2), 274–287. http://doi.org/10.1016/j.respol.2006.11.005

Arthur, W. B., & Polak, W. (2006). The Evolution of Technology with a Simple Computer Model. *Complexity*, *11*(5), 23–31. http://doi.org/10.1002/cplx

Auerswald, P., Kau, S., Lobo, H., & Shell, K. (2000). The production recipes approach to modeling technological innovation: An application to learning by doing. *Journal of Economic Dynamics & Control*, *24*, 389–450.

Axtell, R. L., Casstevens, R., Hendrey, M., Kennedy, W., & Litsch, W. (2013). *Competitive Innovation and the Emergence of Technological Epochs Classification: Social Sciences Short title: Competitive Innovation Author contributions:* Retrieved from http://www.css.gmu.edu/~axtell/Rob/Research/Pages/Technology_files/Tech Epochs.pdf

Baker, N.R., Siegman J., R. A. H. (1967). The Effects of Perceived Needs and Means on the Generation of Ideas for Industrial Research and Development Projects. *IEEE Transactions on Engineering Management*, (December).

Baldwin, C. Y., & Clark, K. B. (2006). Between "Knowledge" and "The Economy": The Notes on the Scientific Study of Designs. In B. Kahin & D. Foray (Eds.), *Advancing Knowledge and The Knowledge Economy* (pp. 298–328). Cambridge, MA: The MIT Press.

Baldwin, Carliss Y., Clark, K. B. (2000). *Design Rules: The Power of Modularity*. Cambridge, MA: MIT Press.

Barenblatt, G. I. (1996). *Scaling, Self-similarity, and Intermediate Asymptotics: Dimensional Analysis and Intermediate Asymptotics*. New York, New York, USA: Cambridge University Press.

Basnet, S., & Magee, C. L. (2016). *Dependence of technological improvement on artifact interactions*. Retrieved from http://arxiv.org/abs/1601.02677

Benson, C. L., & Magee, C. L. (2015a). Quantitative Determination of Technological Improvement from Patent Data. *PloS One*, (April). http://doi.org/DOI:10.1371/journal.pone.0121635 April 15, 2015

Benson, C. L., & Magee, C. L. (2015b). Technology structural implications from the extension of a patent search method. *Scientometrics*, *102*(3), 1965–1985. http://doi.org/10.1007/s11192-014-1493-2

Braha, D., & Reich, Y. (2003). Topological structures for modeling engineering design processes. *Research in Engineering Design*, *14*(4), 185–199. http://doi.org/10.1007/s00163-003-0035-3

Cameron, P. J. (1995). *Combinatorics: Topics, Techniques, Algorithms* (1st ed.). New York,

New York, USA: Cambridge University Press.

Carter, C.F. and Williams, B.R. (1959). *Carter, C.F. and Williams, B.R., 1959. Investment in Innovation*. (London: Oxford University Press.

Carter, C.F., Williams, B. R. (1957). *Industry and Technical Progress: Factors Governing the Speed of Application of Science to Industry*. London: Oxford University Press.

Christensen, B. T., & Schunn, C. D. (2007). The relationship of analogical distance to analogical function and preinventive structure: the case of engineering design. *Memory & Cognition*, *35*(1), 29–38. http://doi.org/10.3758/BF03195939

Christensen, C. M., & Bower, J. L. (1996). Customer Power, Strategic Investment, and the Failure of Leading Firms. *Strategic Management Journal*, *17*(3), 197–218. http://doi.org/10.1002/(SICI)1097-0266(199603)17:3<197::AID-SMJ804>3.0.CO;2-U

Clement, C. a, Mawby, R., & Giles, D. E. (1994). The Effects of Manifest Relational Similarity on Analog Retrieval. *Journal of Memory and Language*. http://doi.org/10.1006/jmla.1994.1019

Dahl, D. W., & Moreau, P. (2002). The Influence and Value of Analogical Thinking During New Product Ideation. *Journal of Marketing Research*, *39*(1), 47–60.

Dasgupta, S. (1996). *Creativity and Technology*. Oxford University Press.

de Solla Price, D. J. (1986). Sealing wax and string. In *Little Science, Big Science and beyond*. New York, New York, USA: Columbia University Press.

Dosi, G. (1982). Technological paradigms and technological trajectories. *Research Policy*, *11*(3), 147–162. http://doi.org/10.1016/0048-7333(82)90016-6

Farmer, J. D., & Lafond, F. (2015). *How predictable is technological progress?* Retrieved from http://arxiv.org/abs/1502.05274

Fehrenbacker, K. (2012). We can thank Moore's Law for the VC cleantech bust. Retrieved from http://gigaom.com/2012/02/01/we-can-thank-moores-law-for-the-vc-cleantech-bust/

Finke, R. A., Ward, T. B., & Smith, S. M. (1996). *Creative Cognition: Theory, Research, and Applications*. Cambridge, MA: MIT Press.

Fleming, L. (2001). Recombinant Uncertainty in Technological Search. *Management Science*, *47*(1), 117–132. http://doi.org/10.1287/mnsc.47.1.117.10671

Fleming, L., & Sorenson, O. (2004). Science as a map in technological search. *Strategic Management Journal*, *25*(89), 909–928. http://doi.org/10.1002/smj.384

Frischknecht, B., Gonzalez, R., Papalambros, P. Y., & Reid, T. (2009). A design science approach to analytical product design. *International Conference on Engineering Design, Design Society, Palo Alto, CA*, (August), 35–46.

Fu, K., Chan, J., Cagan, J., Kotovsky, K., Schunn, C., & Wood, K. (2013). The Meaning of "Near" and "Far": The Impact of Structuring Design Databases and the Effect of Distance of Analogy on Design Output. *Journal of Mechanical Design*, *135*(2), 021007. http://doi.org/10.1115/1.4023158

Gentner, D., & Markman, A. B. (1997). Structure mapping in analogy and similarity.

*American Psychologist*, *52*(1), 45–56.

Gero, J. S., & Kannengiesser, U. (2004). The situated function-behaviour-structure framework. *Design Studies*, *25*(4), 373–391. http://doi.org/10.1016/j.destud.2003.10.010

Girifalco. (1991). *Dynamics of Technological Change*. New York, New York, USA: Van Nostrand Reinhold.

Goel, A. K. (1997). Design, analogy, and creativity. *IEEE Expert*, *12*(3).

Gold, B., The, S., Economics, I., & Sep, N. (1974). Evaluating Scale Economies : The Case of Japanese Blast Furnaces. *The Journal of Industrial Economics*, *23*(1), 1–18.

Gribbin, J. (2002). *The Scientists: A History of Science Told Through the Lives of Its Greatest Inventors*. New York, New York, USA: Random House.

Hatchuel, A., & Weil, B. (2009). C-K design theory: An advanced formulation. *Research in Engineering Design*, *19*(4), 181–192. http://doi.org/10.1007/s00163-008-0043-4

Henderson, R. M., & Clark, K. B. (1990). Architectural Innovation : The Reconfiguration of Existing Product Tech- nologies and the Failure of Established Firms. *Administrative Science Quarterly*, *35*(1), 9–30.

Holyoak, K. J., & Thagard, P. R. (1995). *Mental Leaps: Analogy in Creative Thought*. Cambridge, MA: MIT Press.

Hunt, B. J. (2010). *Pursuing Power and Light*. Baltimore, MD: Johns Hopkins University Press.

Klevorick, A. K., Levin, R. C., Nelson, R. R., & Winter, S. G. (1995). On the sources and significance of interindustry differences in technological opportunities. *Research Policy*, *24*(2), 185–205. http://doi.org/10.1016/0048-7333(93)00762-I

Koestler, A. (1964). *The Act of Creation*. London: Hutchinson & Co.

Koh, H., & Magee, C. L. (2006). A functional approach for studying technological progress: Application to information technology. *Technological Forecasting and Social Change*, *73*(9), 1061–1083. http://doi.org/10.1016/j.techfore.2006.06.001

Koh, H., & Magee, C. L. (2008a). A functional approach for studying technological progress : Extension to energy technology ☆. *Technological Forecasting and Social Change*, *75*, 735–758. http://doi.org/10.1016/j.techfore.2007.05.007

Koh, H., & Magee, C. L. (2008b). A functional approach for studying technological progress: Extension to energy technology. *Technological Forecasting and Social Change*, *75*(6), 735–758. http://doi.org/10.1016/j.techfore.2007.05.007

Langrish J., Gibbons M., Evans W.G., Jevons, F. R. (1972). *Wealth from Knowledge: A Study of Innovation in Industry*. New York, New York, USA: Halsted/ John Wiley.

Leclercq, P., & Heylighen, A. (2002). Analogies Per Hour. In J. S. Gero (Ed.), *Artificial Intelligence in Design'02* (pp. 285–303). Dordrecht: Kluwar Academic Publishers.

Lienhard, J. H. (2008). *How Invention Begins: Echoes of Old Voices in the Rise of New Machines*. New York, New York, USA: The Oxford University Press, UK.

Linsey, J. S., Markman, A. B., & Wood, K. L. (2012). Design by Analogy: A Study of the WordTree Method for Problem Re-Representation. *Journal of Mechanical Design*, *134*(4).

Linsey, J. S., Wood, K. L., & Markman, A. B. (2008). Modality and representation in analogy. *Artificial Intelligence for Engineering Design, Analysis and Manufacturing*, *22*, 85–100. http://doi.org/10.1017/S0890060408000061

Lipsey, Richard G., Carlaw, Kenneth I., Bekar, C. T. (2006). *Economic Transformations: General Purpose Technologies and Long Term Economic Growth*. New York, New York, USA: The Oxford University Press.

Luo, J., Olechowski, A. L., & Magee, C. L. (2014). Technology-based design and sustainable economic growth. *Technovation*, *34*(11), 663–677. http://doi.org/10.1016/j.technovation.2012.06.005

Magee, C. L., Basnet, S., Funk, J. L., & Benosn, C. L. (2014). *Quantitative empirical trends in technical performance* (No. ESD-WP-2014-22). Cambridge, MA. Retrieved from http://esd.mit.edu/WPS/2014/esd-wp-2014-22.pdf

McNerney, J., Farmer, J. D., Redner, S., & Trancik, J. E. (2011). Role of design complexity in technology improvement. *PNAS*, *108*(38), 9008–9013. http://doi.org/10.1073/pnas.1017298108/-/DCSupplemental.www.pnas.org/cgi/doi/10.1073/pnas.1017298108

Meyers, S., Marquis, D. . (1969). Successful Industrial innovation. Washington, D.C.: National Science Foundation.

Mokyr, J. (2002). *The Gifts of Athena: Historical Origins of the Knowledge Economy*. Princeton: Princeton University Press.

Moore, G. E. (1965). Cramming more components onto integrated circuits. *Electronics*, *38*(8), 1–4.

Mowery, D., & Rosenberg, N. (1979). The influence of market demand upon innovation: a critical review of some recent empirical studies. *Research Policy*, *8*(2), 102–153. http://doi.org/10.1016/0048-7333(79)90019-2

Musson, A. E. (1972). *Science, technology and economic growth in the eighteenth century*. (A. E. Musson, Ed.) (1st ed.). Routledge.

Musson, A. E., & Robinson, E. (1989). *Science and Technology in the Industrial Revolution*. Gordon and Breach Science Publishers.

Muth, J. F. (1986). Search Theory and the Manufacturing Progress Function. *Management Science*, *32*(8), 948–962. http://doi.org/10.1287/mnsc.32.8.948

Nagy, B., Farmer, J. D., Bui, Q. M., & Trancik, J. E. (2013). Statistical basis for predicting technological progress. *PloS One*, *8*(2), e52669. http://doi.org/10.1371/journal.pone.0052669

Nelson, Richard R., Winter, S. G. (1982). *An Evolutionary Theory of Economic Change*. Cambridge, MA: Harvard University Press.

Nemet, G., Johnson, E. (2012). Do important inventions benefit from knowledge originating in other technological domains? *Research Policy*, *41*(1).

Nordhaus, W. D. (1996). Do Real-Output and Real-Wage Measures Capture Reality? The History of Lighting Suggests Not. In *The Economics of New Goods* (pp. 27–70). Retrieved from http://www.nber.org/chapters/c6064.pdf

Polanyi, M. (1962). *Personal Knowledge: Towards a Post-Critical Philosophy*. Chicago, IL: University of Chicago Press.

Polya, G. (1945). *How to Solve It: A New Aspect of Mathematical Method* (1st ed.). Princeton, NJ: Princeton University Press.

Popper, K. (1959). *Logic of Scientific Discovery* (1st ed.). Hutchinson & Co.

Romer, P. M. (1990). Endogenous Technological Change. *Journal of Political Economy*, *98*(5).

Rosenberg, N. (1982). *Inside the Black Box: Technology and Economics*. Cambridge, MA: Cambridge University Press.

Rosenberg, N., & Birdzell, L. E., J. (1986). *How the West Grew Rich: The Economic Transformation of the Industrial World*. US: Basic Books.

Ruttan, V. W. (1959). Usher and Schumpeter on Invention , Innovation , and Technological Change Author ( s ): Vernon W . Ruttan Reviewed work ( s ): Published by : Oxford University Press. *The Quarterley Journal of Economics*, *73*(4), 596–606.

Ruttan, V. W. (2001). *Technology, Growth, and Development: An Induced Innovation Perspective*. New York, New York, USA: Oxford University Press.

Sahal, D. (1979). A Theory of Progress Functions. *AIIE Transactions*, *11*(1), 23–29. Retrieved from http://scholar.google.com/scholar?hl=en&btnG=Search&q=intitle:A+I+I+E+Transactions#8

Sahal, D. (1985). Technological guideposts and innovation avenues. *Research Policy*, *14*(2), 61–82. http://doi.org/10.1016/0048-7333(85)90015-0

Schofield, R. (1963). *The Lunar Society of Birmingham: A Social History of Provincial Science and Industry in Eighteenth-Century England*. Clarendon.

Schumpeter, J. A. (1934). *The Theory of Economic Development*. Cambridge, MA: Harvard University Press.

Shai, O., Reich, Y., & Rubin, D. (2009). Creative conceptual design: Extending the scope by infused design. *CAD Computer Aided Design*, *41*(3), 117–135. http://doi.org/10.1016/j.cad.2007.11.004

Simon, H. A. (1962). The Architecture of Complexity. *Proceedings of the American Philosophical Society*, *26*(6), 467–482. http://doi.org/10.1016/S0016-0032(38)92229-X

Simon, H. A. (1969). *The Sciences of the Artificial* (1st ed.). Cambridge, MA: The MIT Press.

Simon, H. A. (1996). *The Sciences of the Artificial* (3rd ed.). Cambridge, MA: The MIT Press.

Solow, R. M. (1956). A Contribution to the Theory of Economic Growth. *The Quarterly Journal of Economics*, *70*(1), 65–94. http://doi.org/10.2307/1884513

Suh, N. P. (2001). *Axiomatic Design: Advances and Applications* (1st ed.). New York, New York, USA: The Oxford University Press, UK.

Taguchi, G. (1992). *Taguchi on Robust Technology Development: Bringing Quality Engineering Upstream*. Asme Press Series.

Toynbee, A. J. (1962). Introduction: The Geneses of Civilizations. In *A Study of History, 12 Vol*. New York, New York, USA.

Tseng, I., Moss, J., Cagan, J., & Kotovsky, K. (2008). The role of timing and analogical similarity in the stimulation of idea generation in design. *Design Studies*, *29*(3), 203–221. http://doi.org/10.1016/j.destud.2008.01.003

Tushman, M. L., & Anderson, P. (1986). Technological Discontinuities and Organizational Environments life cycles. *Administrative Science Quarterly*, *31*, 439–465.

Usher, A. P. (1954). *A History of Mechanical Inventions* (1st ed.). New York, New York, USA: Beacon Press, Beacon Hill, MA.

Utterback, J. M. (1974). Innovation in industry and the diffusion of technology. *Science (New York, N.Y.)*, *183*(4125), 620–626. http://doi.org/10.1126/science.183.4125.620

Vincenti, W. (1990). *What Engineers Know, and How They Know It*. Baltimore, MD: John Hopkins University Press.

Weber, C., & Deubel, T. (2003). NEW THEORY-BASED CONCEPTS FOR PDM AND PLM Property-Driven Development / Design ( PDD ), 1–10.

Weisberg, R. W. (2006). Creativity. In *Creativity* (1st ed., pp. 153–2007). Hoboken, NJ: John Wiley & Sons, Inc.

Whitney, D. E. (1996). Why Mechanical Design Will Never be Like VLSI design. *Research in Engineering Design*, *8*, 125–138.

Whitney, D. E. (2004). Physical limits to modularity. In *MIT Engineering System Division Internal Symposium*. Retrieved from https://esd.mit.edu/symposium/pdfs/papers/whitney.pdf

Wright, T. P. (1936). Factors Affecting the Cost of Airplanes. *Journal of Aero. Science*, 122–138.

Yelle, L. E. (2007). The learning curve: historical review and comprehensive survey. *Decision Sciences*.

Youn, H., Bettencourt, L. M. a., Strumsky, D., & Lobo, J. (2014). Invention as a Combinatorial Process: Evidence from U.S. Patents. *Physics Society*, *June*, 1–22. Retrieved from http://arxiv.org/abs/1406.2938v1